\documentclass[]{aa} 

\usepackage{natbib}
\bibpunct{(}{)}{;}{a}{}{,} % to follow the A&A style
\usepackage{epsfig}

\newcommand{\beq}{\begin{equation}} 
\newcommand{\eeq}{\end{equation}} 
\newcommand{\beqa}{\begin{eqnarray}} 
\newcommand{\eeqa}{\end{eqnarray}} 
\newcommand{\bea}{\begin{array}} 
\newcommand{\ea}{\end{array}} 

\newcommand{\dd}{{\rm d}}

\newcommand{\lag}{\langle} 
\newcommand{\rag}{\rangle} 

\newcommand{\ii}{{\rm i}}

\newcommand{\vk}{{\bf k}}

\newcommand{\tkappa}{{\tilde{\kappa}}}

\newcommand{\tW}{{\tilde{W}}}

\newcommand{\vtheta}{\vec{\theta}}
\newcommand{\vell}{\vec{\ell}}

\newcommand{\cD}{{\cal{D}}}

\newcommand{\Pkappa}{P_{\kappa}}
\newcommand{\Bkappa}{B_{\kappa}}
\newcommand{\xikappa}{\xi_{\kappa}}
\newcommand{\zetakappa}{\zeta_{\kappa}}

\newcommand{\Map}{M_{\rm ap}}
 
\begin{document} 

\topmargin =0.1cm

\title{Modeling of weak-lensing statistics. II. Configuration-space statistics.}    
\author{
Patrick Valageas\inst{1}
\and
Masanori Sato\inst{2}
\and
Takahiro Nishimichi\inst{3}
}   
\institute{
Institut de Physique Th\'eorique, CEA Saclay, 91191 Gif-sur-Yvette, France 
\and
Department of Physics, Nagoya University, Nagoya 464-8602, Japan
\and
Institute for the Physics and Mathematics of the Universe, University of
Tokyo, Kashiwa, Chiba 277-8568, Japan
}
\date{Received / Accepted } 
 
\abstract
{}
{We investigate the performance of an analytic model of the 3D matter distribution,
which combines perturbation theory with halo models, for weak-lensing
configuration-space statistics.}
{We compared our predictions for the weak-lensing convergence two-point
and three-point correlation functions with numerical simulations and fitting formulas
proposed in previous works. We also considered the second- and third-order moments
of the smoothed convergence and of the aperture-mass.}
{As in our previous study of Fourier-space weak-lensing statistics, we find
that our model agrees better with simulations than previously
published fitting formulas. Moreover, we recover the dependence
on cosmology of these weak-lensing statistics and we can describe multi-scale
moments. This approach allows us to
obtain the quantitative relationship between these integrated weak-lensing
statistics and the various contributions to the underlying 3D density
fluctuations, decomposed over perturbative, two-halo, or one-halo terms.}
{}


\maketitle

\section{Introduction} 
\label{Introduction}

The standard paradigm known as the $\Lambda$CDM cosmology includes
dark components: dark matter and dark energy \citep{Komatsu2011}.
Weak lensing of background galaxies by foreground large-scale structures,
the so-called ``cosmic shear'',  has been recognized as a
potentially powerful tool for probing the distribution of dark matter as
well as the nature of dark energy~\citep{Albrecht2006}.
Reports of significant detections of cosmic shear have been made by
various groups \citep{Bacon2000,VanWaerbeke2000,Wittman2000,Hamana2003,Jarvis2006,Semboloni2006,Fu2008,Schrabback2010}.

By analyzing the cosmic shear data, one can directly measure the
power spectrum of the matter density fluctuations on cosmological scales, 
which contain a wealth of cosmological information such as neutrino masses 
and dark energy equation-of-state parameters~\citep{Jarvis2006,Semboloni2006,Ichiki2009,Schrabback2010}.
Consequently, it is a main goal of cosmology to infer and constrain these quantities
from observations.
To do this, a number of ambitious surveys are planned, such as the Hyper
Suprime-Cam Weak Lensing
Survey~\citep{Miyazaki2006}\footnote{http://www.naoj.org/Projects/HSC/index.html},
the Dark Energy Survey~(DES)\footnote{http://www.darkenergysurvey.org/},
the Large Synoptic Survey Telescope~(LSST)\footnote{http://www.lsst.org/},  
Euclid~\citep{Refregier2010}\footnote{http://www.euclid-ec.org/}, and the Wide-Field Infrared Survey
Telescope~(WFIRST)\footnote{http://wfirst.gsfc.nasa.gov/}.

Most weak-lensing information is contained in small angular scales
and therefore weak-lensing statistics are nonlinear and non-Gaussian~\citep{Munshi2008,Takada2009,Sato2009,Sato2011a}.
If we aim to exploit the full information, we have to treat the
nonlinear effects to accurately model the weak lensing statistics.
Furthermore, one has to use an appropriate likelihood function with given
marginal distributions, otherwise the obtained results would be
systematically biased \citep{Sato2010,Sato2011}.

In a first companion paper \citep[][hereafter paper I]{Valageas2011f}, we
studied the Fourier-space weak-lensing statistics such as the weak-lensing
power spectrum and bispectrum, and found that our
model proposed by \citet{Valageas2011d,Valageas2011e},
which combines perturbation theory with halo models,
agrees better with ray-tracing simulations than previously published
fitting formulas and phenomenological models.
In this second paper, we study the real-space weak-lensing statistics, which
are more often used for the statistical analysis of actual measurements than
Fourier-space statistics, because observations are made in configuration space.

Previous works have already shown that on small scales the halo model provides
a good description of the two-, three- and four-point correlations or smoothed
moments of the cosmic shear (using some approximations)
\citep{Takada2002,Takada2003,Benabed2006},
whereas a stochastic halo model can recover the probability distribution function
of the unsmoothed convergence \citep{Kainulainen2011,Kainulainen2011a}.
Here we include all ``one-halo'', ``two-halo'' and ``three-halo" terms, as well as
one-loop perturbative results, and we compare these with larger-scale simulations.
This yields a greater accuracy and allows us to compare these
different contributions, from very large to small scales.
This should be useful for practical purposes because
these different terms have different theoretical accuracies and probe different
regimes of gravitational clustering, hence it is important to know their relative
impact on weak-lensing probes as a function of angular scale.

This paper is organized as follows.
In Sect.~\ref{Analytic} we briefly recall how configuration-space weak-lensing
statistics are computed from polyspectra of the 3D matter density field. We describe
our numerical simulations and the data analysis
in Sect.~\ref{Numerical}. Then, we present detailed comparisons between the
simulation results, previous models, and our
theoretical predictions for two-point functions in Sect.~\ref{Lensing-two-point}
and three-point functions in Sect.~\ref{Lensing-three-point}.
We study the relative importance of the different contributions arising from ``one-halo'',
``two-halo'',  or ``three-halo'' terms in Sect.~\ref{contributions}.
Then, we check the robustness of our model when we vary the cosmological 
parameters in Sect.~\ref{Cosmology} and we briefly study multi-scale
moments in Sect.~\ref{multi-scale}.
Finally, we conclude in Sect.~\ref{Conclusion}.

\section{From 3D statistics to weak-lensing statistics}
\label{Analytic}

\subsection{Lensing power spectrum and bispectrum}
\label{lensing-power}

Using Born's approximation, the weak-lensing convergence $\kappa(\vtheta)$ can be
written as the integral of the density contrast along the line of sight
\citep{Bartelmann2001,Munshi2008},
\beq
\kappa(\vtheta) = \int_0^{\chi_s} \dd\chi \, w(\chi,\chi_s) \, \delta(\chi,\cD\vtheta) ,
\label{kappa-def}
\eeq
where $\chi$ and $\cD$ are the radial and angular comoving distances,
\beq
w(\chi,\chi_s) = \frac{3\Omega_{\rm m} H_0^2 \cD(\chi) \cD(\chi_s-\chi)}{2c^2\cD(\chi_s)} (1+z) ,
\label{w-def}
\eeq
and $z_s$ is the redshift of the source (in this article we only consider the case where
all sources are located at a single redshift to simplify the comparisons with numerical
simulations and the dependence on the source redshift).
Then, using a flat-sky approximation, which is valid for small angles below a few
degrees \citep{Valageas2011c}, we define its 2D Fourier transform through
\beq
\kappa(\vtheta) = \int \dd\vell \, e^{\ii\vell\cdot\vtheta} \, \tkappa(\vell) .
\label{kappa-l-def}
\eeq
As in paper I, we define the 2D convergence power spectrum and bispectrum as
\beq
\lag \tkappa(\vell_1) \tkappa(\vell_2) \rag = \delta_D(\vell_1+\vell_2) \Pkappa(\ell_1) ,
\label{Pkappa-def}
\eeq
and
\beq
\lag \tkappa(\vell_1) \tkappa(\vell_2) \tkappa(\vell_3) \rag =
\delta_D(\vell_1+\vell_2+\vell_3) \, \Bkappa(\ell_1,\ell_2,\ell_3) .
\label{Bkappa-def}
\eeq
From Eq.(\ref{kappa-def}) one obtains at once from Limber's approximation
\citep{Limber1953,Kaiser1992,Bartelmann2001,Munshi2008}
\beq
\Pkappa(\ell)= 2\pi \int_0^{\chi_s} \dd\chi \, \frac{w^2}{\cD^2} \, P(\ell/\cD;z) ,
\label{Pkappa-P}
\eeq
\beq
\Bkappa(\ell_1,\ell_2,\ell_3)= (2\pi)^2 \!\! \int_0^{\chi_s} \!\! \dd\chi \, 
\frac{w^3}{\cD^4} \, B(\ell_1/\cD,\ell_2/\cD,\ell_3/\cD;z) ,
\label{Bkappa-B}
\eeq
where $P(k;z)$ and $B(k_1,k_2,k_3;z)$ are the 3D power spectrum and bispectrum
of the matter density contrast at redshift $z$.
As described in paper I, this provides the weak-lensing convergence power spectrum
and bispectrum from the model we developed in \cite{Valageas2011d,Valageas2011e}
for the 3D power spectrum and bispectrum through a simple integration over the radial
coordinate up to the source plane.

\subsection{Configuration-space statistics}
\label{real-space}

We focus here on configuration-space weak-lensing statistics, which may be more
convenient than Fourier-space quantities for practical purposes because of complex
survey geometries.
Indeed, observations of the shear field are made in configuration space, by
measuring large-scale correlations of galaxy ellipticities, and going to Fourier space
(with further operations that are not local in real space) can be difficult because
galaxy surveys do not cover the whole sky and shear maps can show irregular
boundaries and internal holes due to observational constraints.

In particular, we consider the two-point and three-point correlation functions of the
weak-lensing convergence, defined as
\beq
\lag \kappa(\vtheta_1)\kappa(\vtheta_2) \rag = \xikappa(|\vtheta_2-\vtheta_1|) ,
\label{xi-def}
\eeq
\beq
\lag \kappa(\vtheta_1)\kappa(\vtheta_2)\kappa(\vtheta_3) \rag = \zetakappa(|\vtheta_2-\vtheta_3|,|\vtheta_3-\vtheta_1|,|\vtheta_1-\vtheta_2|) ,
\label{zeta-def}
\eeq
where we used statistical homogeneity and isotropy (hence $\zetakappa$ only 
depends on the lengths of the three sides of the triangle defined by the summits
$\{\vtheta_1,\vtheta_2,\vtheta_3\}$ and the bispectrum (\ref{Bkappa-B})
only depends on the three lengths $\{\ell_1,\ell_2,\ell_3\}$).

These real-space correlations can be expressed in terms of the Fourier-space power 
spectrum and bispectrum as
\beq
\xikappa(\theta) = 2\pi \int_0^{\infty} \dd\ell \, \ell \, \Pkappa(\ell) \,
J_0(\ell \theta),
\label{xi-P}
\eeq
where $\theta=|\vtheta_2-\vtheta_1|$ is the pair angular distance as in
Eq.(\ref{xi-def}), and
\beqa
\zetakappa(\nu_1,\nu_2,\nu_3) & = & \frac{\pi}{\nu_1^2\nu_2^2} \int_0^{\infty}
\dd\ell \, \ell^3 \int_0^{\pi/2} \dd\theta \, \sin(2\theta) \int_0^{2\pi} 
\dd\varphi \nonumber \\
&& \hspace{-1.5cm} \times \Bkappa(\ell_1,\ell_2,\ell_3) \, 
J_0(\ell\sqrt{1+\sin(2\theta)\cos(\varphi+\alpha_3)}) ,
\label{zeta-B}
\eeqa
where $\{\nu_1,\nu_2,\nu_3\}$ are the lengths of the three sides of the triangle
$\{\vtheta_1,\vtheta_2,\vtheta_3\}$ as in Eq.(\ref{zeta-def}), $\alpha_3$
is the inner angle at summit $\vtheta_3$, we chose without loss of generality
\beq
\nu_1 \geq \nu_2 \geq \nu_3 , \;\;\; \cos\alpha_3 =
\frac{\nu_1^2+\nu_2^2-\nu_3^2}{2\nu_1\nu_2} ,
\label{nu-order}
\eeq
and we noted in Eq.(\ref{zeta-B}) the multipoles
\beq
\ell_1= \frac{\ell\cos\theta}{\nu_2} , \;\; \ell_2= \frac{\ell\sin\theta}{\nu_1} ,
\;\; \ell_3^2= \ell_1^2+\ell_2^2+2\ell_1\ell_2\cos\varphi .
\label{l1-l2-l3}
\eeq

In addition to the three-point correlation $\zetakappa$ of the convergence, it can
be useful to consider the three-point correlation of the cosmic shear ${\vec\gamma}$.
Because the latter is a spin-2 quantity, one is led to consider several three-point
shear correlations (or ``natural components''), depending on the choice of the reference
direction (or of the projection procedure of the cosmic shear vectors), see
\citet{Schneider2003,Schneider2005}. However, they can all be written as integrals
over the convergence bispectrum, such as Eq.(\ref{zeta-B}), or integrals over
the convergence three-point correlation (\ref{zeta-def}) \citep{Shi2011}.
Therefore, although
we only consider the convergence three-point correlation (\ref{zeta-def}) in this paper,
we can expect a similar level of agreement between our analytical model and simulations
for these other three-point correlations (in addition we also consider the third-order
moment of the aperture-mass, which can be related to both the convergence and
the cosmic shear).

It is also common practice to study smoothed averages $X_s$ of the convergence
or shear field, defined by their filtering window $W^{X_s}_{\theta_s}(\vtheta)$ through
\beq
X_s = \int \frac{\dd\vtheta}{\pi\theta_s^2} \,
\kappa(\vtheta) \, W^{X_s}_{\theta_s}(\vtheta) .
\label{X-def}
\eeq
For instance, the smoothed convergence $\kappa_s$ is defined by a top-hat filtering,
\beq
W^{\kappa_s}_{\theta_s}(\vtheta) = 1 \;\; \mbox{if} \;\; |\vtheta|<\theta_s 
\;\; \mbox{and zero otherwise} ,
\label{W-kappas-def}
\eeq
while the ``aperture-mass'' $\Map$ is defined by a compensated filter 
\citep{Schneider1996,VanWaerbeke2001}, such as
\beq
W^{\Map}_{\theta_s}(\vtheta) = 3  \left(1-\frac{\theta^2}{\theta_s^2}\right)
\left(1-3\frac{\theta^2}{\theta_s^2}\right)  \;\; \mbox{if} \;\; 
|\vtheta|<\theta_s ,
\label{W-Map-def}
\eeq
and $W^{\Map}_{\theta_s}(\vtheta) = 0$ if $|\vtheta|>\theta_s$.
This also reads in Fourier space as
\beq
X_s = \int \dd\vell \, \tkappa(\vell) \, \tW^{X_s}_{\theta_s}(\vell\theta_s)
\eeq
with
\beq
\tW^{X_s}_{\theta_s}(\vell\theta_s) = \int \frac{\dd\vtheta}{\pi\theta_s^2}
\, e^{\ii\vell\cdot\vtheta} \, W^{X_s}_{\theta_s}(\vtheta) .
\eeq
In particular, we have 
\beq
\tW^{\kappa_s}_{\theta_s}(\ell\theta_s) = 2 
\frac{J_1(\ell\theta_s)}{\ell\theta_s} , \;\;\;
\tW^{\Map}_{\theta_s}(\ell\theta_s) = 24 
\frac{J_4(\ell\theta_s)}{(\ell\theta_s)^2} .
\label{tW-kappa-Map}
\eeq

We mainly focus here on one-point moments of $\kappa_s$ and $\Map$, that is
$\lag X_s^p \rag$, and we do not consider multi-point statistics such as
$\lag X_s(\vtheta_1;\theta_{s1}) .. X_s(\vtheta_p;\theta_{sp})\rag$ associated
with $p$ windows centered on $p$ different directions and with $p$ different angular
radii.
However, we will check the validity of our model for multi-scale statistics,
that is, for windows of different sizes centered on the same direction, 
in Sect.~\ref{multi-scale}.

In the simpler case of one-point moments $\lag X_s^p \rag$
the variance reads as
\beq
\lag X_s^2\rag = 2\pi \int_0^{\infty} \dd\ell \, \ell \, \Pkappa(\ell) \,
\tW^{X_s}_{\theta_s}(\ell \theta_s)^2 ,
\label{Xs-variance}
\eeq
while the third-order moment reads as
\beqa
\lag X_s^3\rag & = & 24\pi \int_0^{\infty}\dd\ell_1 \, \ell_1 \int_{\ell_1/2}^{\ell_1}
\dd\ell_2 \, \ell_2 \int_0^{\arccos[\ell_1/(2\ell_2)]} \dd\varphi 
\nonumber \\
&& \hspace{-0.5cm} \times \Bkappa(\ell_1,\ell_2,\ell_3) \, 
\tW^{X_s}_{\theta_s}(\ell_1 \theta_s) \tW^{X_s}_{\theta_s}(\ell_2 \theta_s) 
\tW^{X_s}_{\theta_s}(\ell_3 \theta_s) ,
\label{Xs-skewness}
\eeqa
where we used the symmetries of the bispectrum and we noted
\beq
\ell_3^2= \ell_1^2+\ell_2^2 - 2\ell_1\ell_2\cos\varphi .
\eeq
In practice, to avoid the numerous oscillations and changes of sign brought by
the Fourier-space filters $\tW^{X_s}_{\theta_s}$ given in
Eq.(\ref{tW-kappa-Map}), we found it convenient to express the third-order
moment (\ref{Xs-skewness}) in terms of the real-space three-point correlation
(\ref{zeta-def}), although this yields a five-dimensional integral instead of the
three-dimensional integral (\ref{Xs-skewness}),
\beqa
\lag X_s^3\rag & = & \frac{24\pi}{(\pi\theta_s^2)^3} \int_0^{\theta_s} \!\!
\dd\theta_1 \theta_1 W^{X_s}_{\theta_s}(\theta_1) \int_0^{\theta_1} \!\!
\dd\theta_2 \theta_2 W^{X_s}_{\theta_s}(\theta_2) \nonumber \\
&& \hspace{-0.5cm} \times \int_0^{\theta_2} \!\! \dd\theta_3 \theta_3 
W^{X_s}_{\theta_s}(\theta_3) \int_0^{\pi} \!\!\! \dd\varphi_2
\int_0^{2\pi} \!\!\! \dd \varphi_3 \, \zetakappa(\nu_1,\nu_2,\nu_3) ,
\label{Xs-3-real}
\eeqa
with
\beqa
\nu_1^2 & = & (\theta_3\cos\varphi_3-\theta_2\cos\varphi_2)^2 + 
(\theta_3\sin\varphi_3-\theta_2\sin\varphi_2)^2 , \\
\nu_2^2 & = &  (\theta_3\cos\varphi_3-\theta_1)^2 
+ \theta_3^2 \sin^2\varphi_3 , \\
\nu_3^2 & = &  (\theta_2\cos\varphi_2-\theta_1)^2 
+ \theta_2^2 \sin^2\varphi_2 .
\eeqa

\section{Numerical simulations}
\label{Numerical}

We performed the ray-tracing simulations through high-resolution $N$-body
simulations of cosmological structure formation
\citep{Jain2000,Hamana2001a,Sato2009,Takahashi2011} to obtain accurate
predictions of the configuration statistics for weak lensing.
To run the $N$-body simulations, we used a modified version of the {\tt
Gadget-2} code~\citep{Springel2005} and employed 256$^3$ particles for
each simulation.
The ray-tracing simulations were constructed from $2\times 200$
realizations of $N$-body simulations with cubic 240 and 480$h^{-1}$Mpc
on a side, respectively, to cover a light cone of angular size
5$^{\circ}\times 5^{\circ}$ \citep[see Fig.~1 in][]{Sato2009}.
For our fiducial cosmology, we adopted the standard $\Lambda$CDM cosmology
with matter fraction $\Omega_{\rm m}=0.238$, baryon fraction
$\Omega_{\rm b}=0.0416$, dark energy fraction $\Omega_{\rm de}=0.762$ with
the equation of state parameters $w_0=-1$ and $w_a=0$, spectral index
$n_s=0.958$, normalization $A_{\rm s}=2.35\times 10^{-9}$, and Hubble
parameter $h=0.732$, which are consistent with the WMAP three-year results~\citep{Spergel2007}. 
This fiducial cosmology gives the normalization $\sigma_8=0.759$ for the
variance of the linear density fluctuations in a sphere of radius 8$h^{-1}$Mpc.
We considered source redshifts at either $z_s=0.6$, 1.0, or 1.5.
Using ray-tracing simulations we generated 1000 realizations of
convergence maps for each source redshift.

In addition to the fiducial cosmology case, we performed
ray-tracing simulations for several slightly different cosmologies.
We varied each of the following cosmological parameters:
$A_s$, $n_s$, the cold dark matter density $\Omega_{\rm c}h^2$,
$\Omega_{\rm de}$, and $w_0$ by $\pm 10\%$, respectively.
Therefore $h$, $\Omega_{\rm m}$, and $\Omega_{\rm b}$ are dependent parameters, because we assumed that the Universe is flat and the baryon
density $\Omega_{\rm b}h^2$ is fixed. For each of these ten different
cosmologies, we obtain 40 realizations of convergence fields for each
of the three source redshifts.
Details of the methods used for the ray-tracing simulations can be found in
\citet{Sato2009}.
All convergence maps used in this paper are the same as those used in paper I.
%but for varied cosmologies we use three orthogonal
%projection axes instead of using only one projection axis to increase
%the independence.

In Sects.~\ref{Lensing-two-point}-\ref{contributions},
we use the maps for the fiducial cosmology, while in
Sect.~\ref{Cosmology} we use those for the varied cosmologies to
investigate the robustness of our model.
In Sect.~\ref{Cosmology} we show the results for six cases,
varying $n_s$, $\Omega_{\rm c}h^2$, and $w_0$ by $\pm 10\%$.
The exact values for these cosmological parameters are listed in
Table A.1 in paper I.

\section{Lensing two-point functions}
\label{Lensing-two-point}

\begin{figure*}
\begin{center}
\epsfxsize=6.1 cm \epsfysize=5.4 cm {\epsfbox{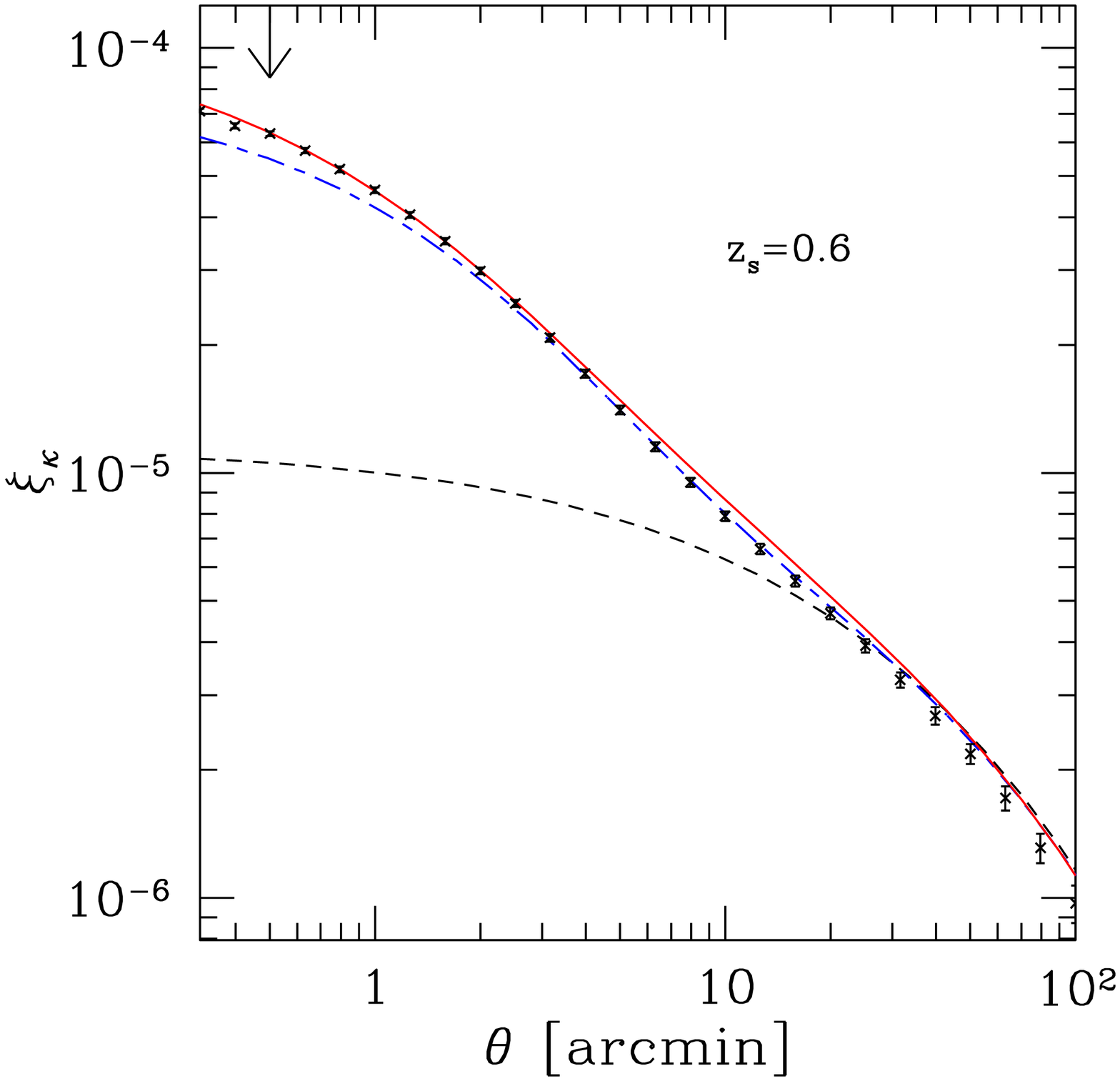}}
\epsfxsize=6.05 cm \epsfysize=5.4 cm {\epsfbox{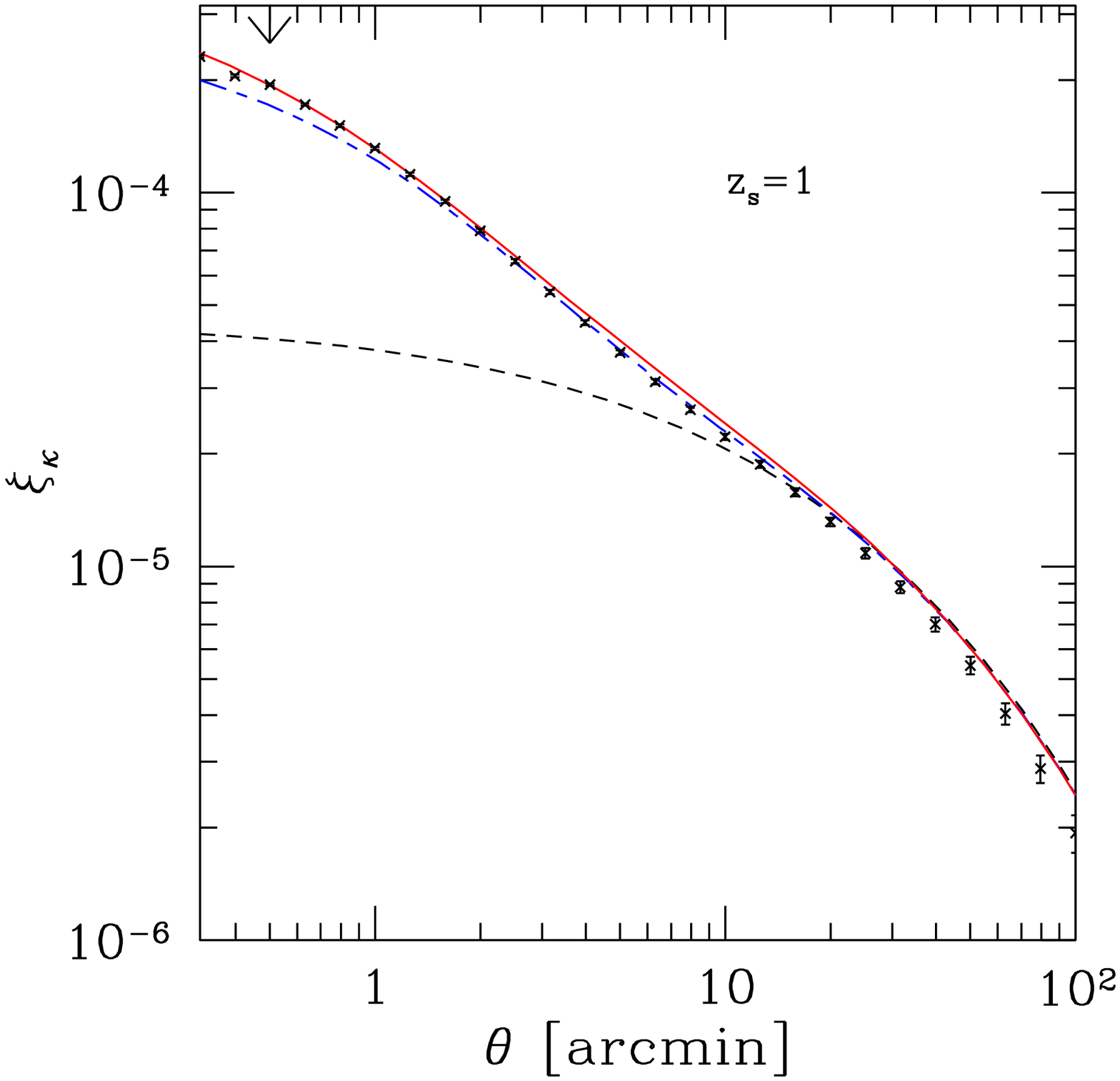}}
\epsfxsize=6.05 cm \epsfysize=5.4 cm {\epsfbox{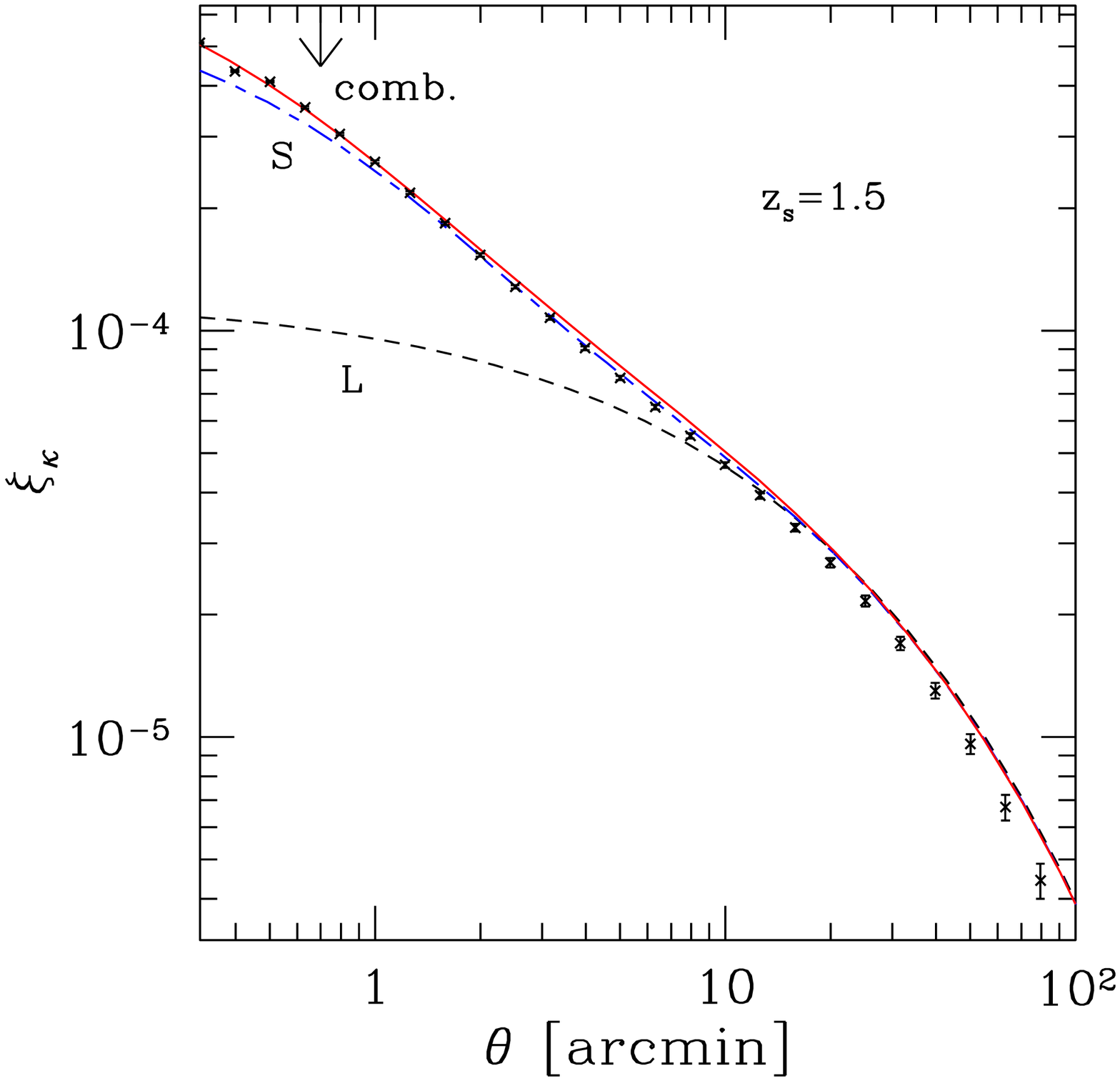}}\\
\epsfxsize=6.1 cm \epsfysize=5.4 cm {\epsfbox{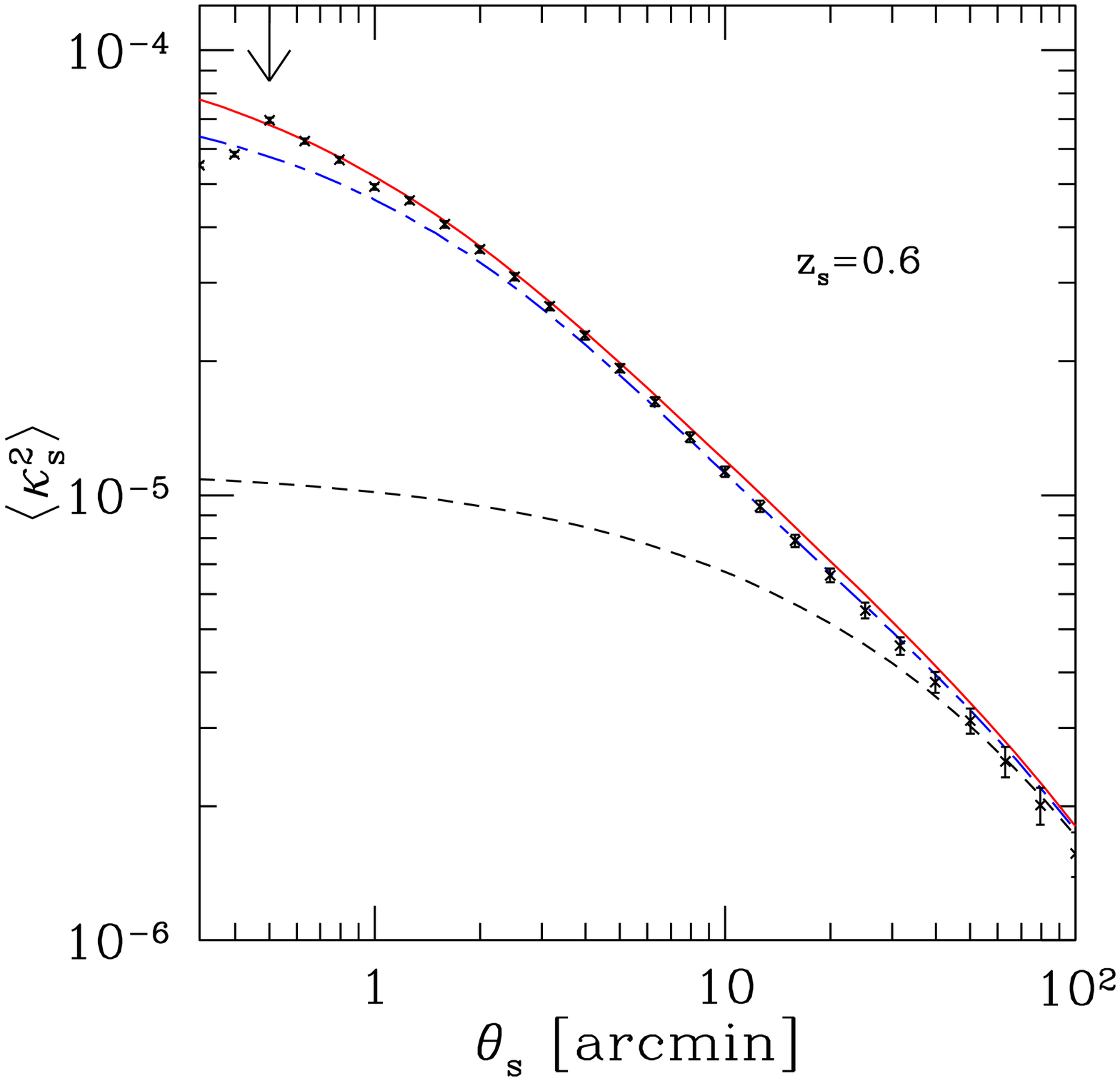}}
\epsfxsize=6.05 cm \epsfysize=5.4 cm {\epsfbox{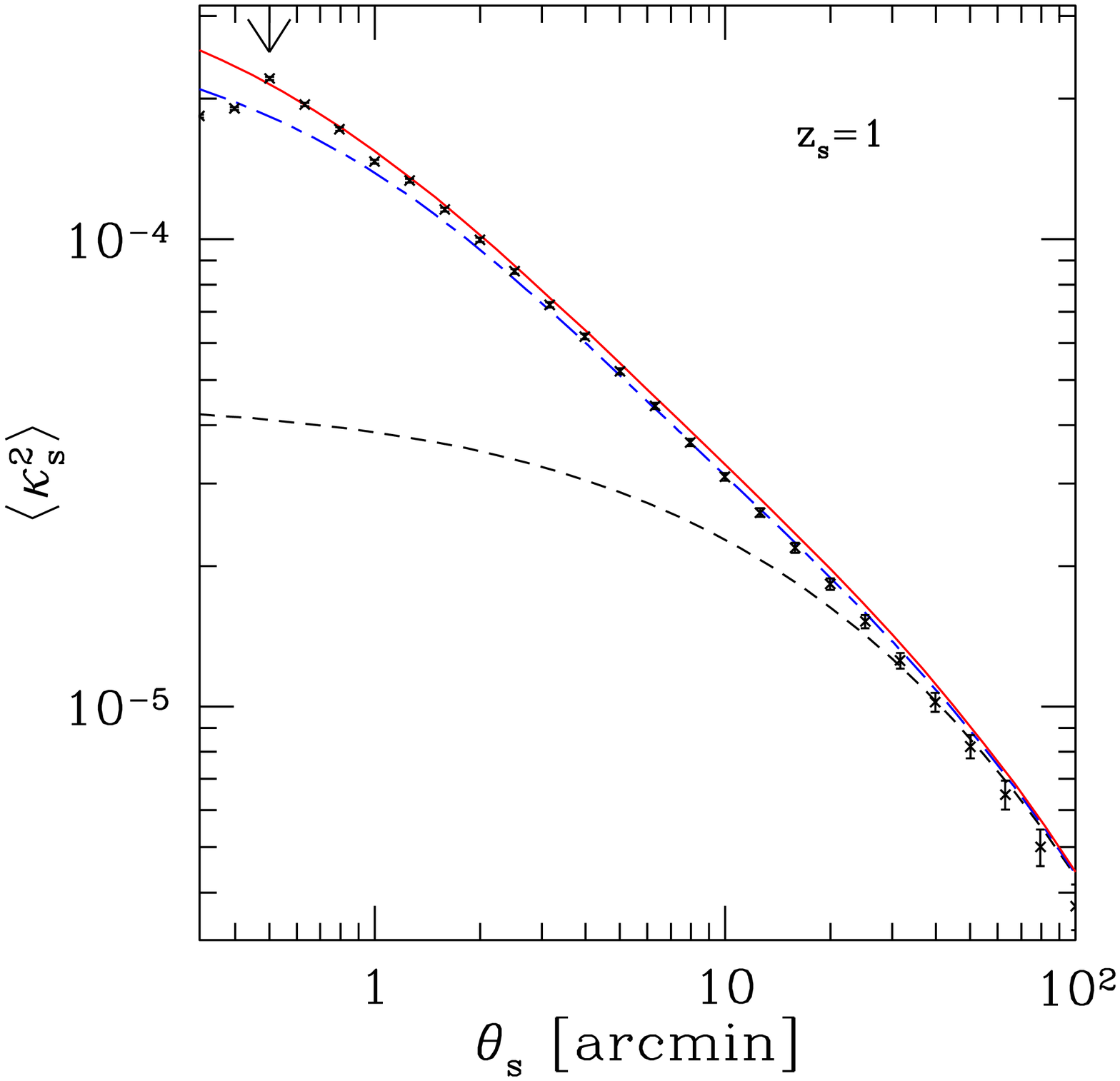}}
\epsfxsize=6.05 cm \epsfysize=5.4 cm {\epsfbox{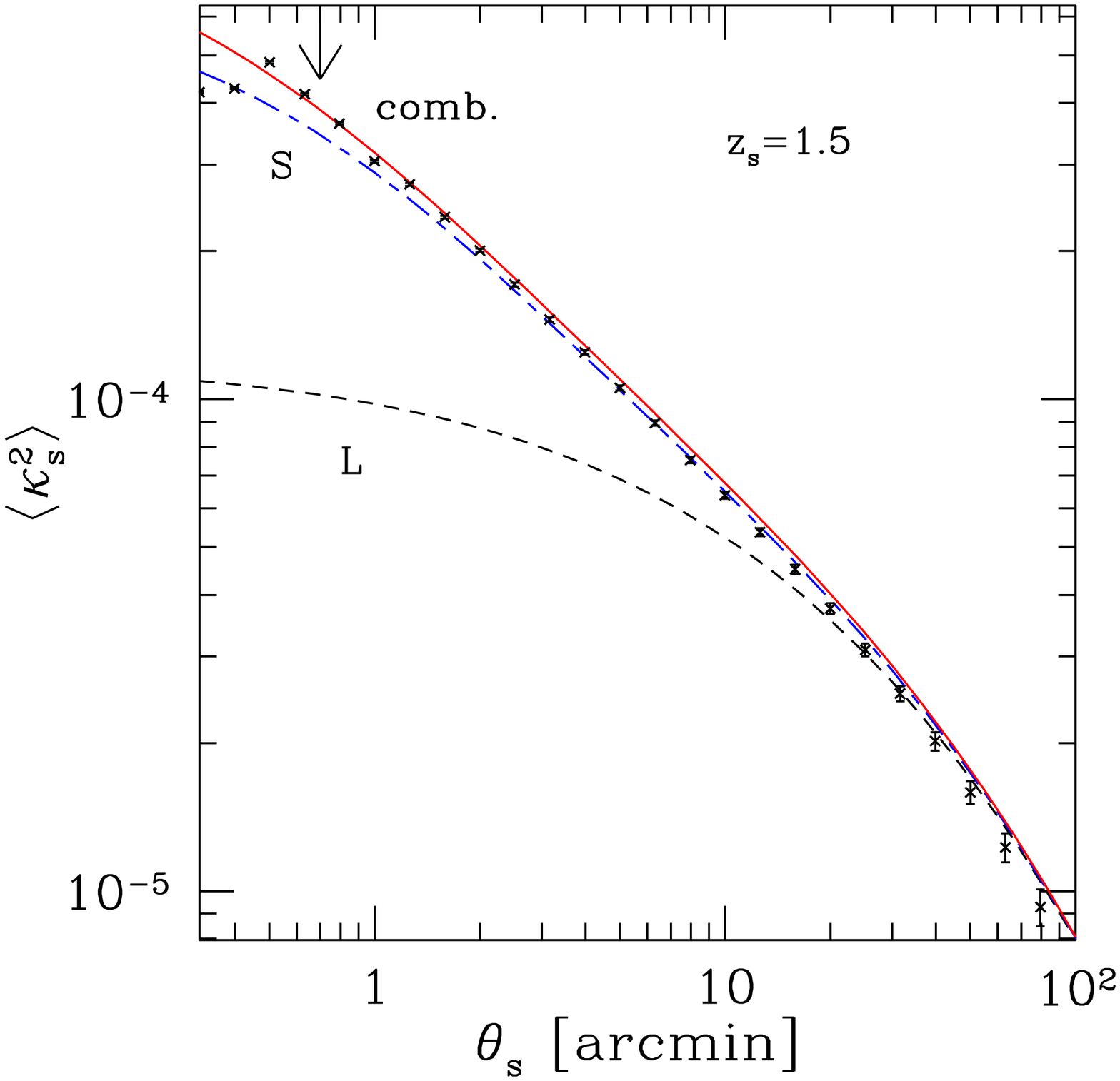}}\\
\epsfxsize=6.1 cm \epsfysize=5.4 cm {\epsfbox{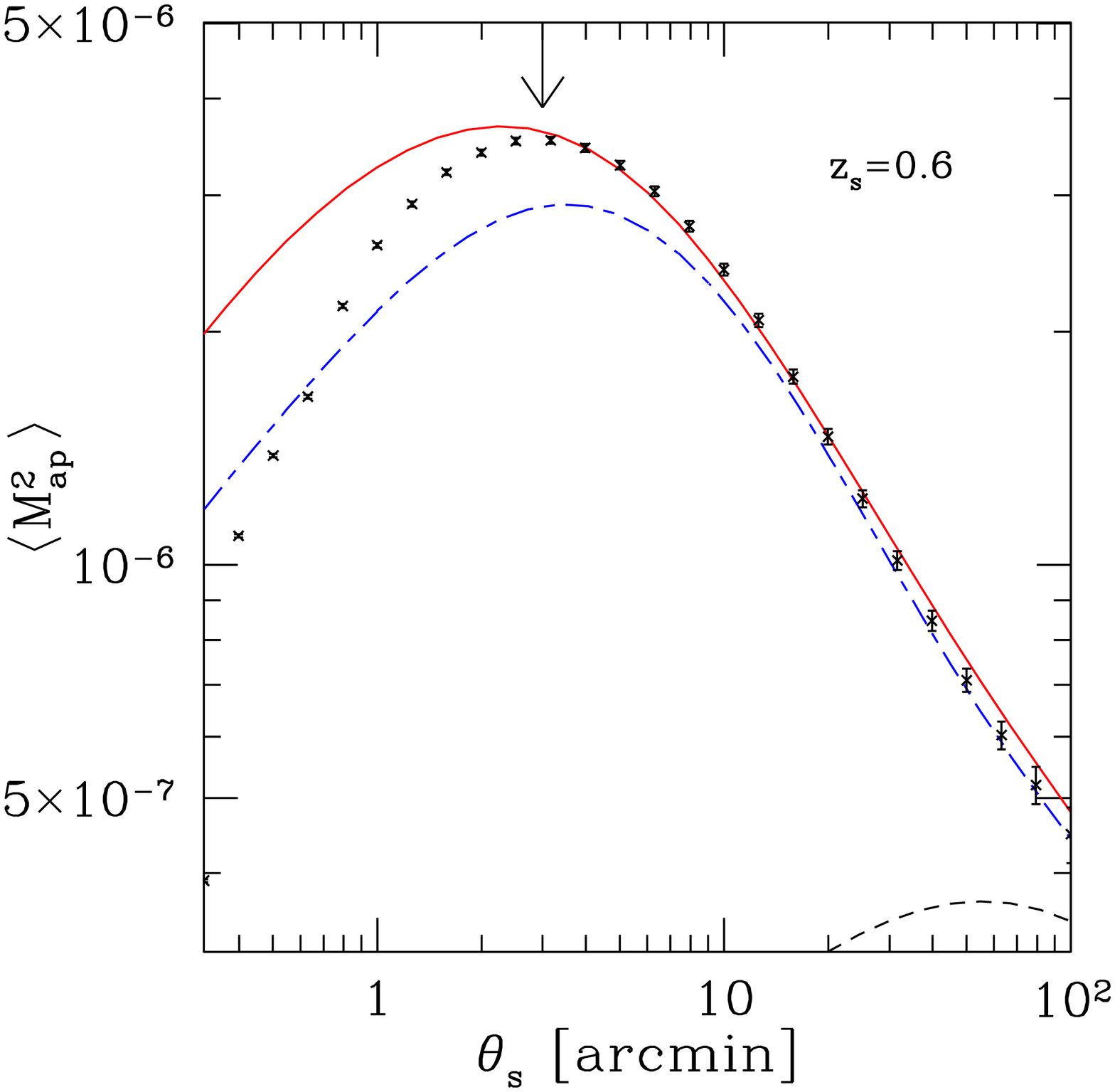}}
\epsfxsize=6.05 cm \epsfysize=5.4 cm {\epsfbox{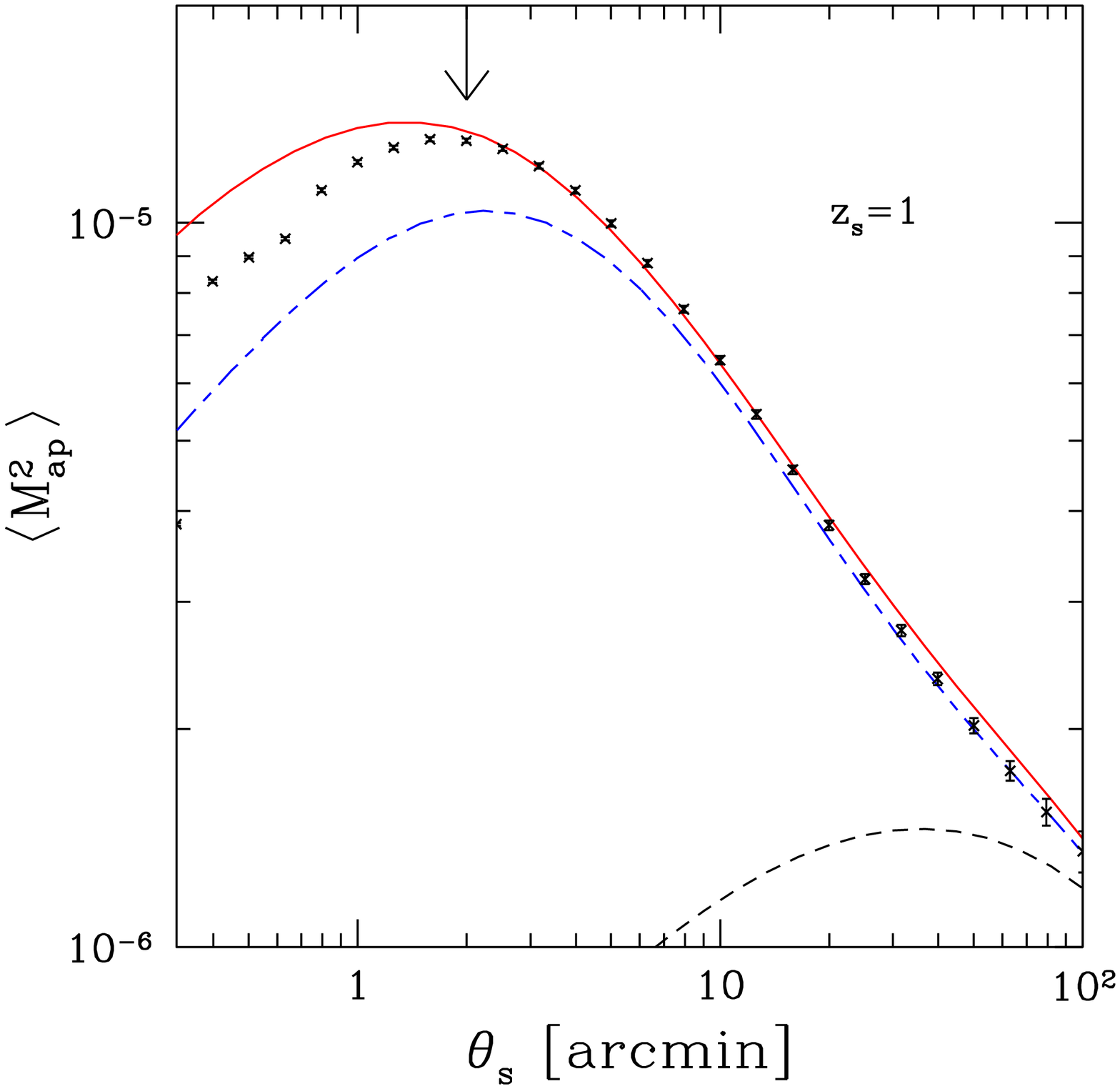}}
\epsfxsize=6.05 cm \epsfysize=5.4 cm {\epsfbox{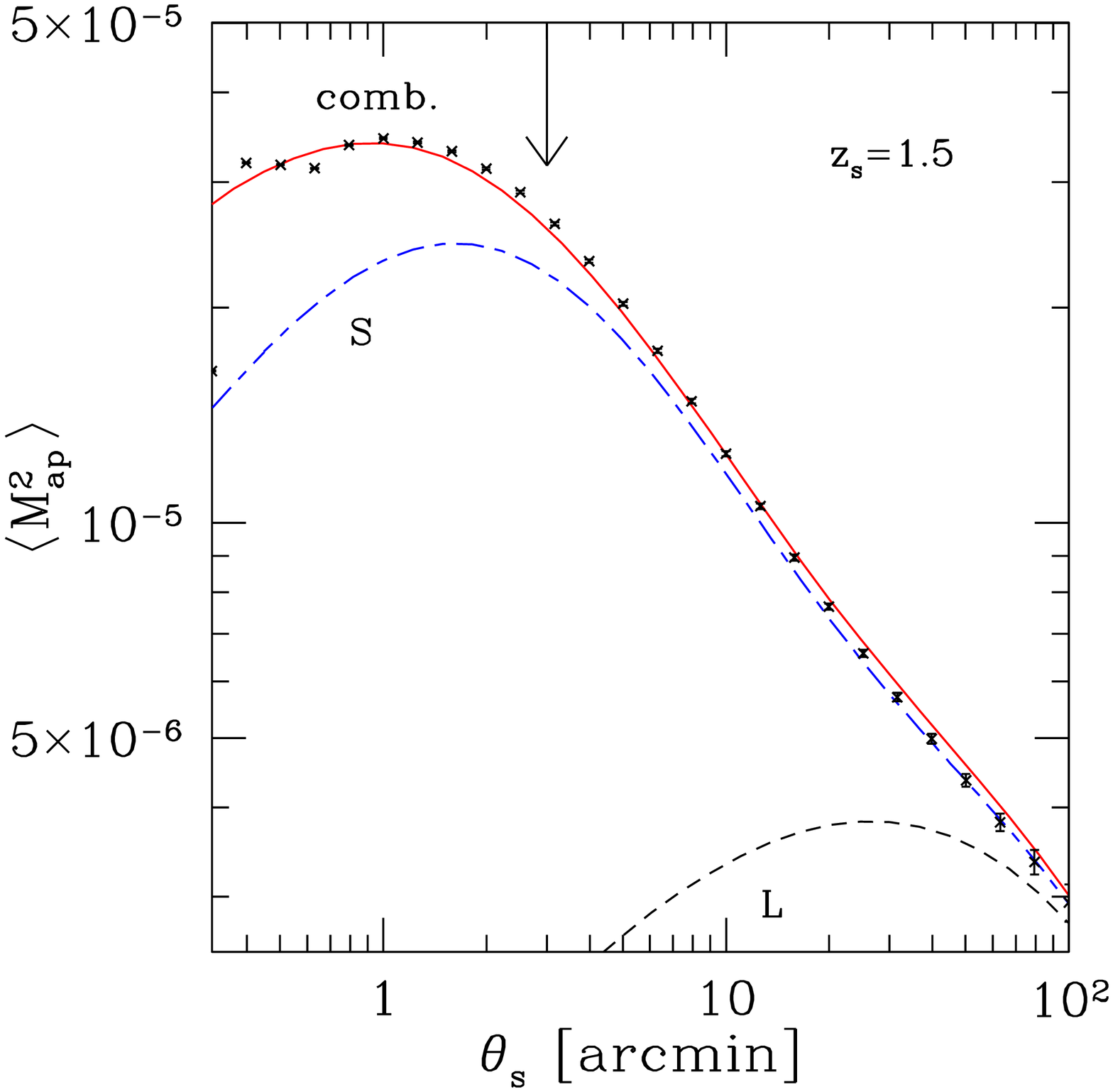}}
\end{center}
\caption{{\it Upper row:} convergence two-point correlation function for sources at
redshifts $z_s=0.6, 1$, and $1.5$, as a function of the angular pair separation $\vtheta$. 
The points are the results from numerical simulations with $3-\sigma$ error bars.
The low black dashed line ``L'' is the linear correlation,
the middle blue dash-dotted line ``S'' is the result from the ``halo-fit''
of \cite{Smith2003}, and the upper red solid line ``comb.'' is the result from
our model, which combines one-loop perturbation theory with a halo model.
The vertical arrow shows the scale down to which the simulation result is valid within
$5\%$.
{\it Middle row:} Variance of the smoothed convergence for the same cases, as a function
of the smoothing angle $\theta_s$.
{\it Lower row:} Variance of the aperture mass for the same cases, as a function
of the smoothing angle $\theta_s$.}
\label{fig_xi}
\end{figure*}

We now compare our results for weak-lensing two-point functions with numerical
simulations. As in paper I, we also considered the predictions obtained from the
popular ``halo-fit'' fitting function for the 3D density power spectrum given
in \cite{Smith2003}, to estimate the advantages of more systematic approaches
like ours.

We show our results for the convergence two-point correlation $\xikappa(\theta)$,
the variance of the smoothed convergence $\lag\kappa_s^2\rag$, and the variance
of the aperture mass $\lag\Map^2\rag$ in Fig.~\ref{fig_xi}.
The numerical error bars increase on large scales because of the finite size of the
simulation box. On small scales the numerical error is dominated by systematic
effects, because of the finite resolution, which leads to an underestimation of the
small-scale
power. This underestimation was clearly apparent  for the power spectrum
$\Pkappa(\ell)$ shown in paper I and can also be seen (especially at low redshift)
for the variance of the aperture mass, which involves a compensated filter and
probes a narrow range of scales \citep{Schneider1996}.
For each source redshift we estimated the angular scale down to which the simulations
have an accuracy of better than $5\%$ by comparing with higher-resolution simulations
(with $512^3$ particles instead of $256^3$). This scale is shown by the vertical
arrow in Fig.~\ref{fig_xi} and we can check that our model indeed agrees
with the numerical simulations down to this angular scale.

The two-point correlation and the smoothed convergence are not as sensitive to this
low-resolution effect because they involve uncompensated filters, $\tW(0)\neq 0$,
which implies that at a given smoothing angular scale $\theta_s$ they receive
greater contributions from larger scales (which are unaffected by the numerical 
resolution) than the aperture mass.

For the same reason, $\xikappa(\theta)$ and $\lag\kappa_s^2\rag$ remain 
adequately described by linear theory down to $\sim 10$ arcmin, whereas
$\lag\Map^2\rag$ already shows significant deviations at $\sim 100$ arcmin.
This also explains why the predictions from our model and the ``halo-fit'' are closer
for the first two statistics than for $\Map$.
In agreement with our results for the convergence power spectrum (paper I) and
previous works \citep{White2004,Hilbert2009,Sato2009,Semboloni2011}, the ``halo-fit'' formula underestimates
the power on small scales. The discrepancy is again larger for the aperture mass.
As in paper I, our model provides a more accurate match to the numerical simulations
down to $\sim 1$ arcmin. On smaller scales the results from simulations show a fast
drop (especially for $\Map$) that is not physical but due to the finite resolution.
This prevents an accurate comparison with our predictions. However, since our model
is built from a physical halo model and has been tested for the 3D density field
down to highly nonlinear scales with higher-resolution simulations 
\citep{Valageas2011d}, it should be more reliable than the numerical results shown
in Fig.~\ref{fig_xi} below $\sim 1$ arcmin.
This shows one advantage of analytic (or semi-analytic) approaches as compared with
numerical simulations: they can provide realistic predictions on a wider
range of scales.

\section{Lensing three-point functions}
\label{Lensing-three-point}

\begin{figure*}
\begin{center}
\epsfxsize=6.1 cm \epsfysize=5.4 cm {\epsfbox{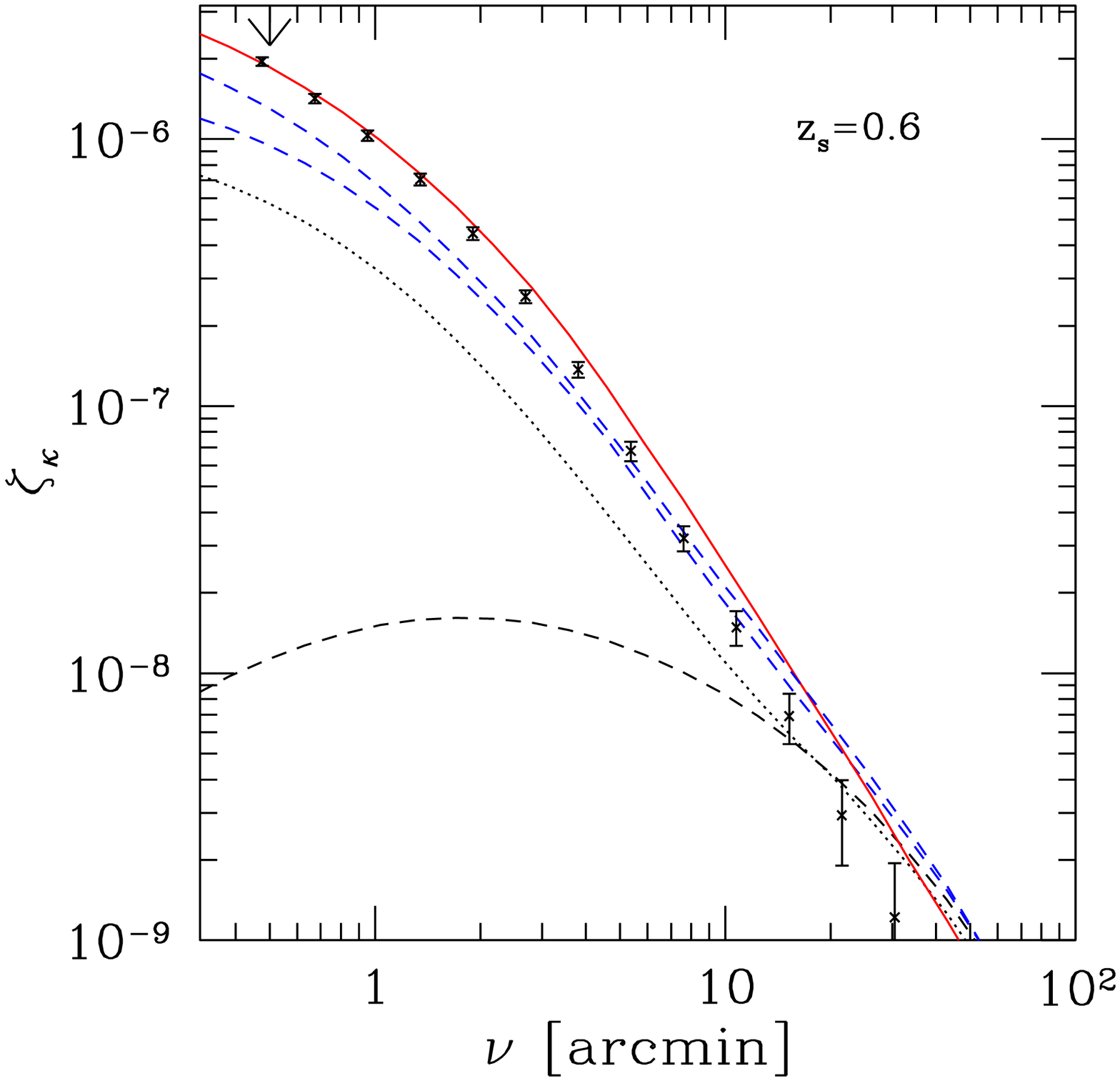}}
\epsfxsize=6.05 cm \epsfysize=5.4 cm {\epsfbox{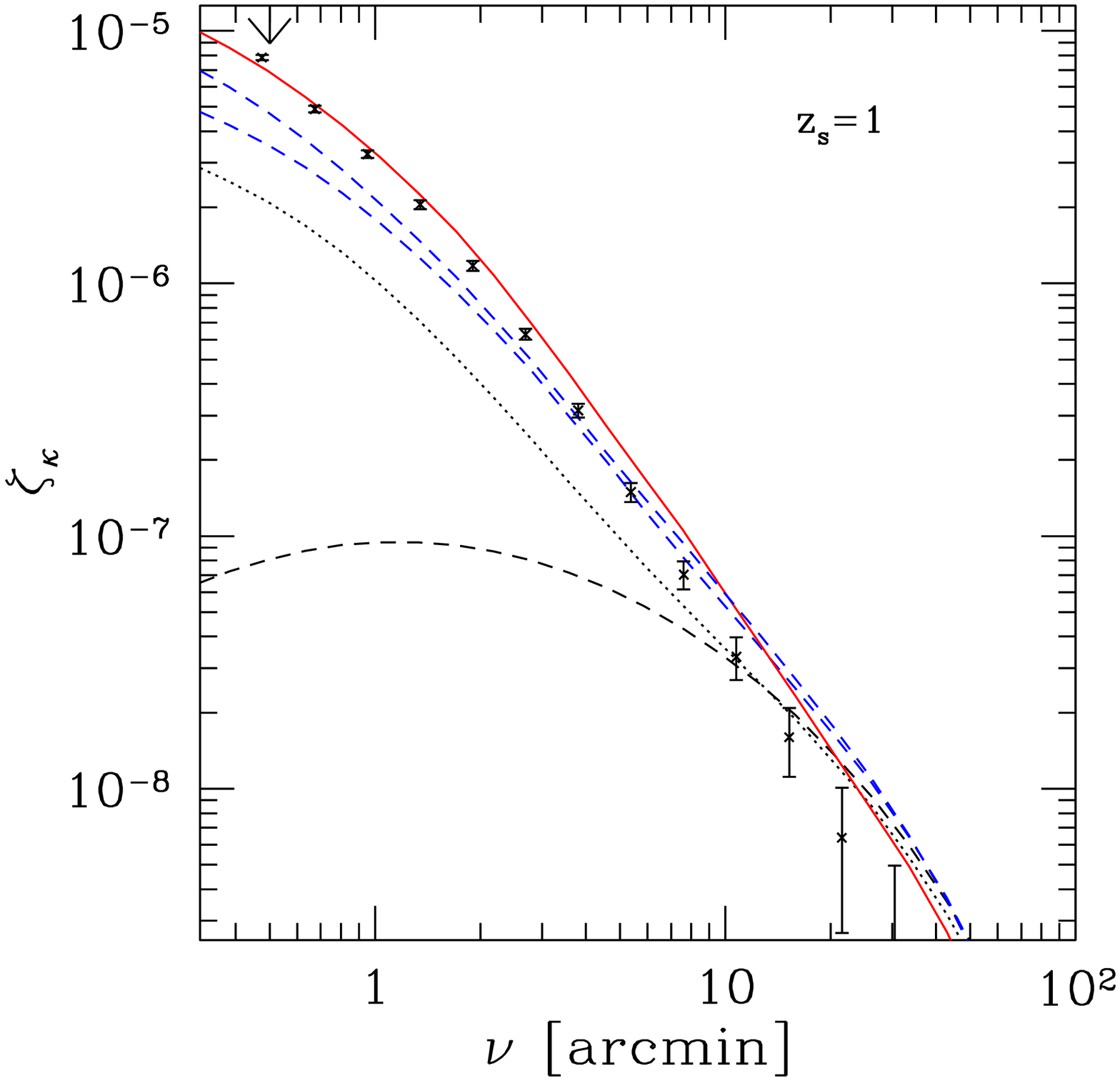}}
\epsfxsize=6.05 cm \epsfysize=5.4 cm {\epsfbox{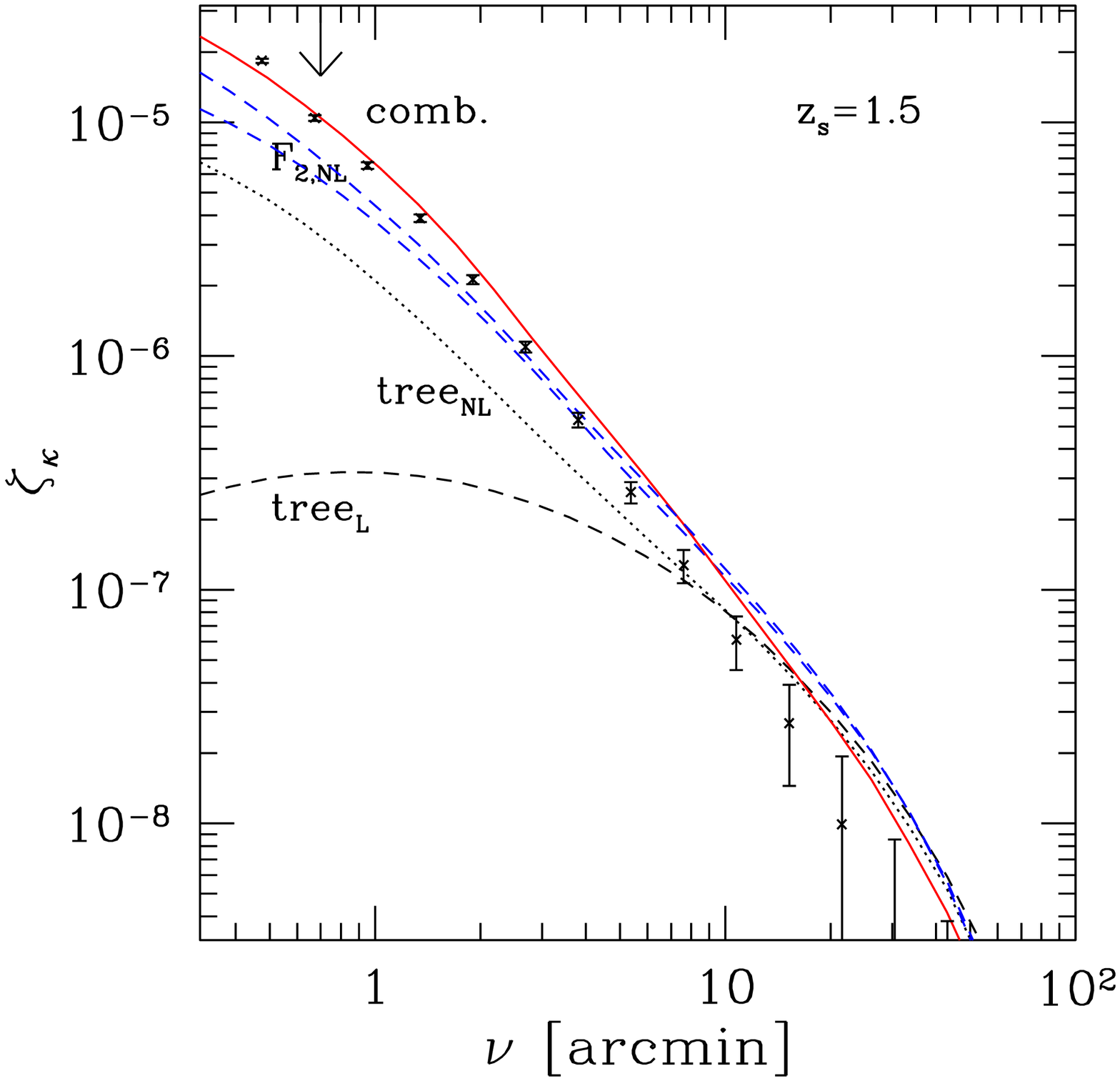}}\\
\epsfxsize=6.1 cm \epsfysize=5.4 cm {\epsfbox{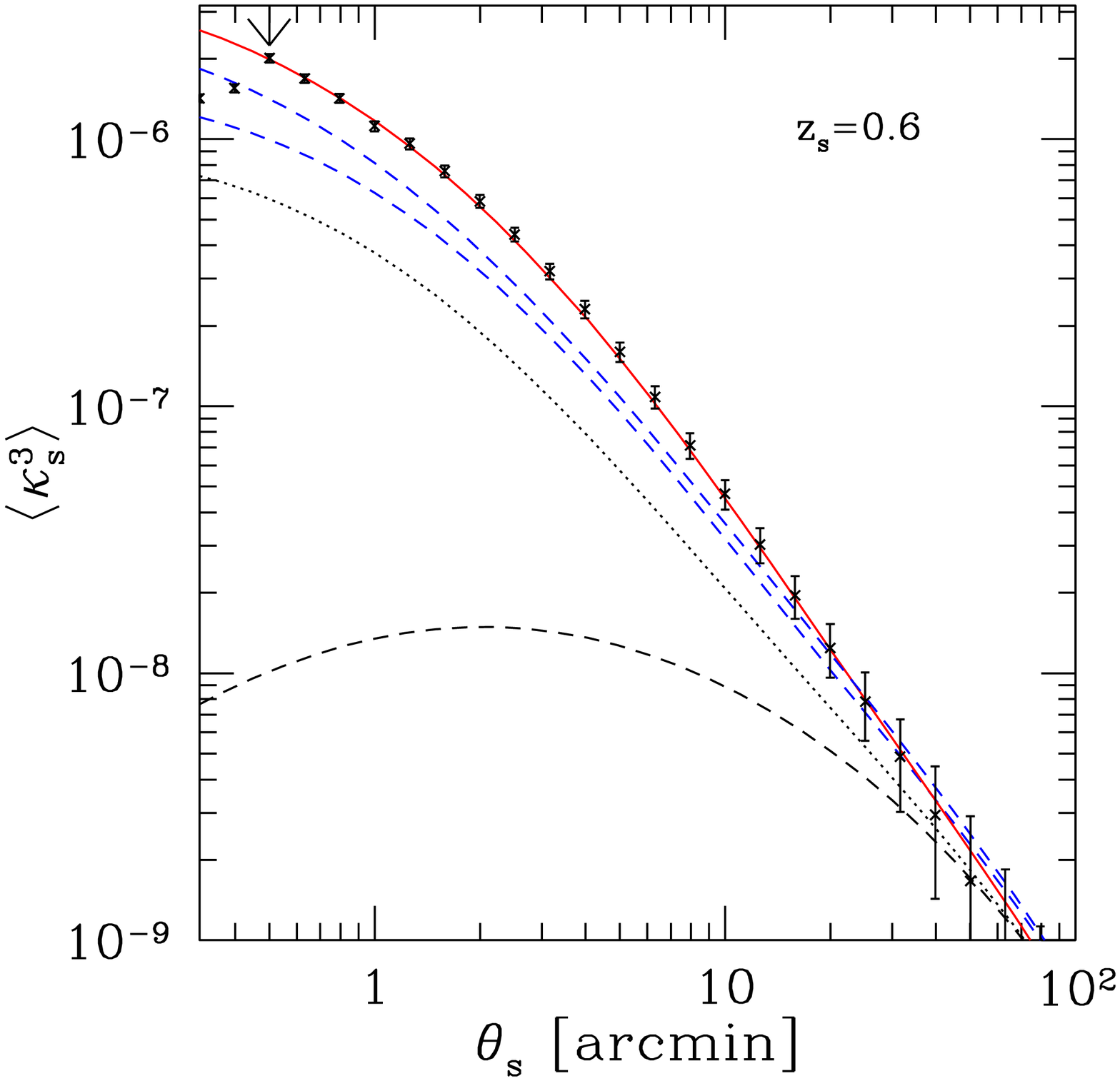}}
\epsfxsize=6.05 cm \epsfysize=5.4 cm {\epsfbox{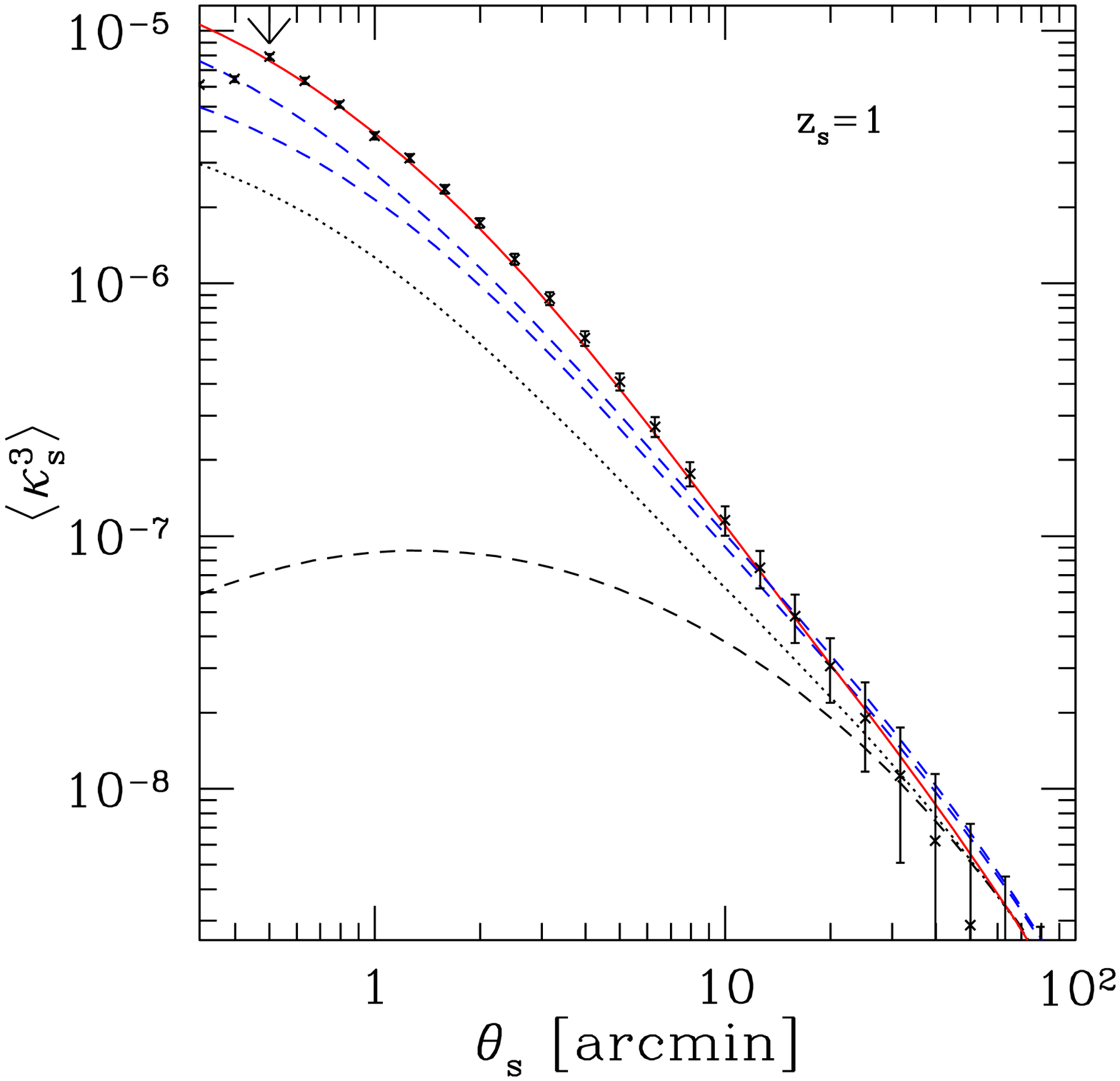}}
\epsfxsize=6.05 cm \epsfysize=5.4 cm {\epsfbox{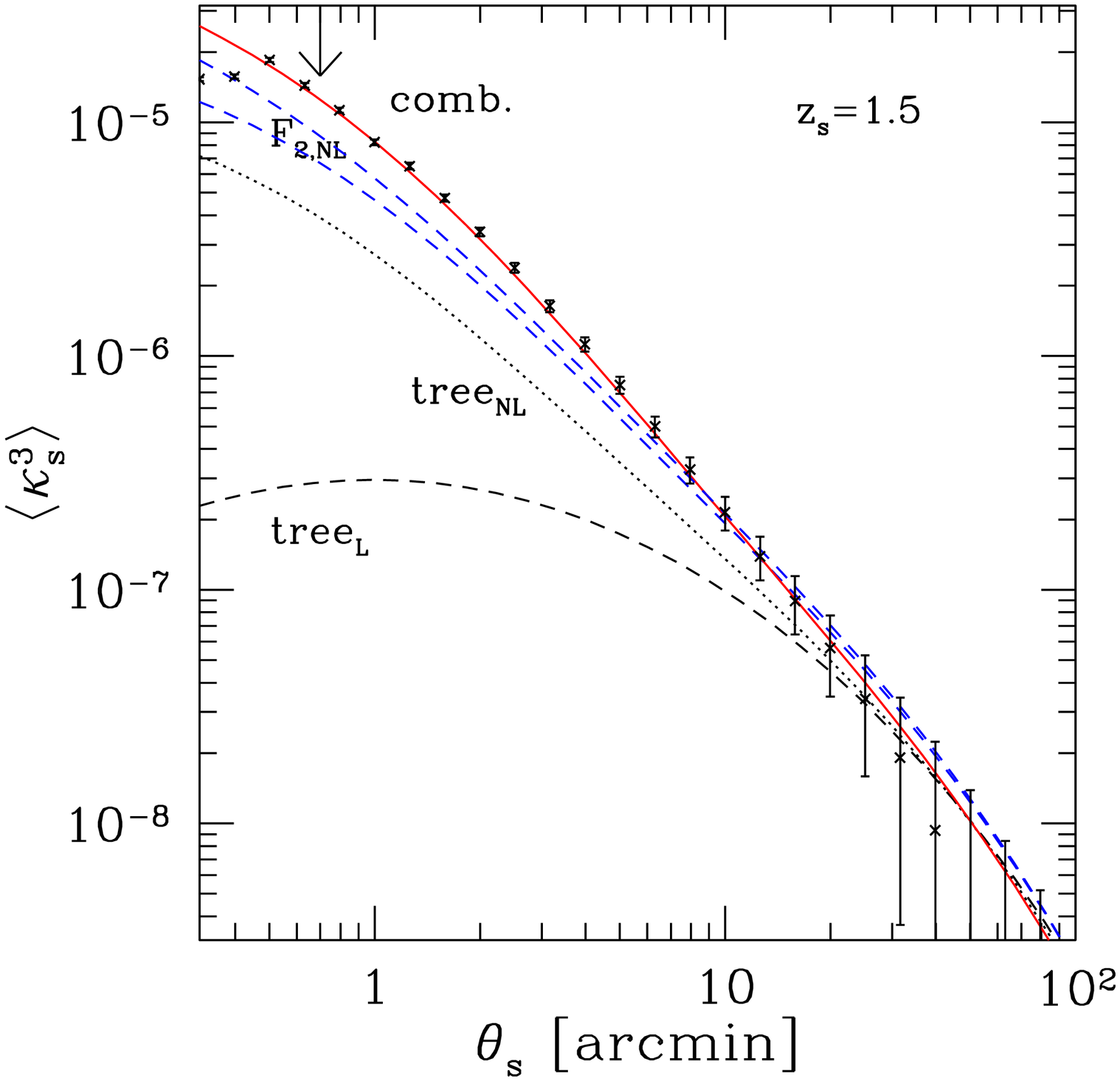}}\\
\epsfxsize=6.1 cm \epsfysize=5.4 cm {\epsfbox{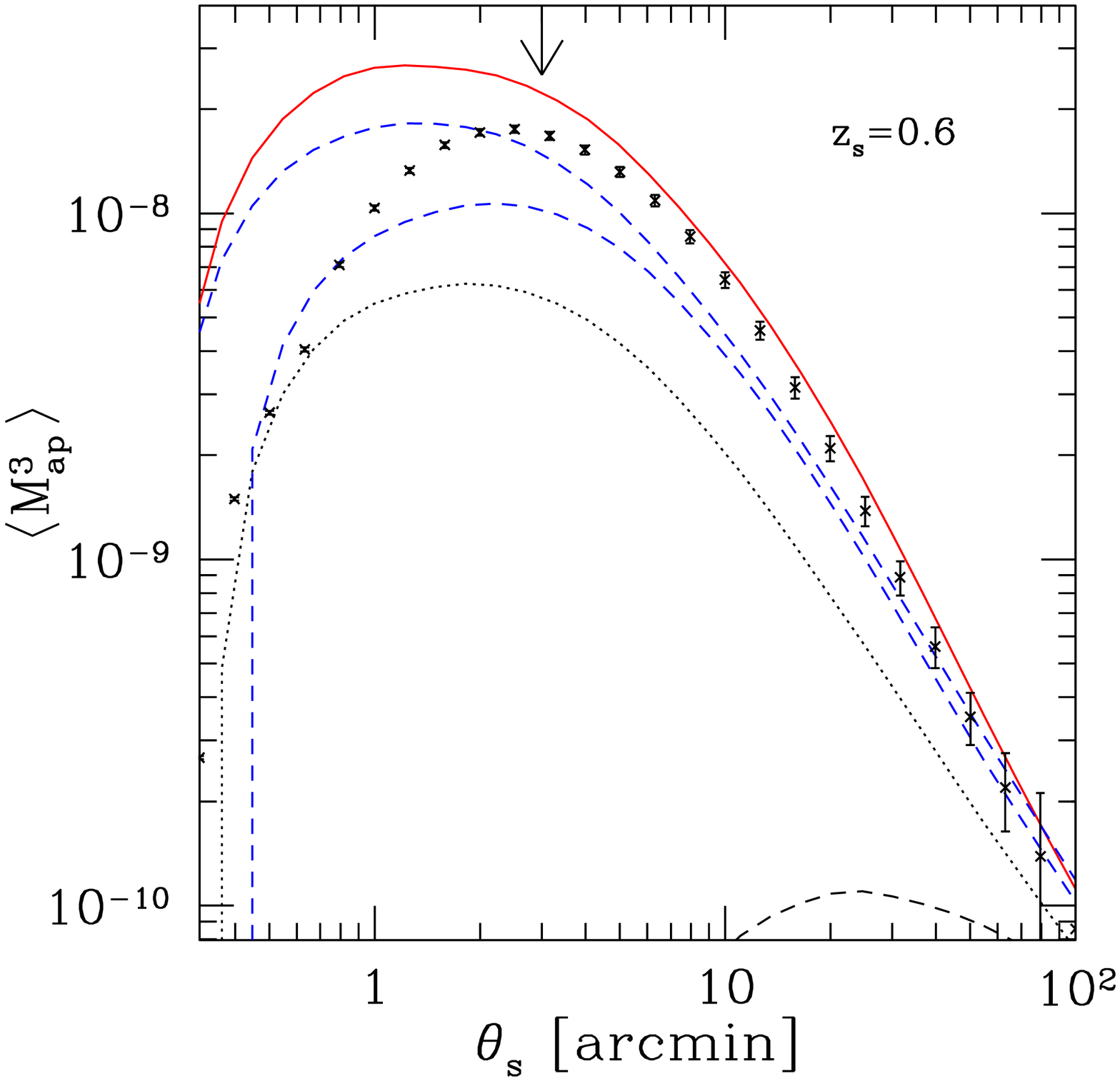}}
\epsfxsize=6.05 cm \epsfysize=5.4 cm {\epsfbox{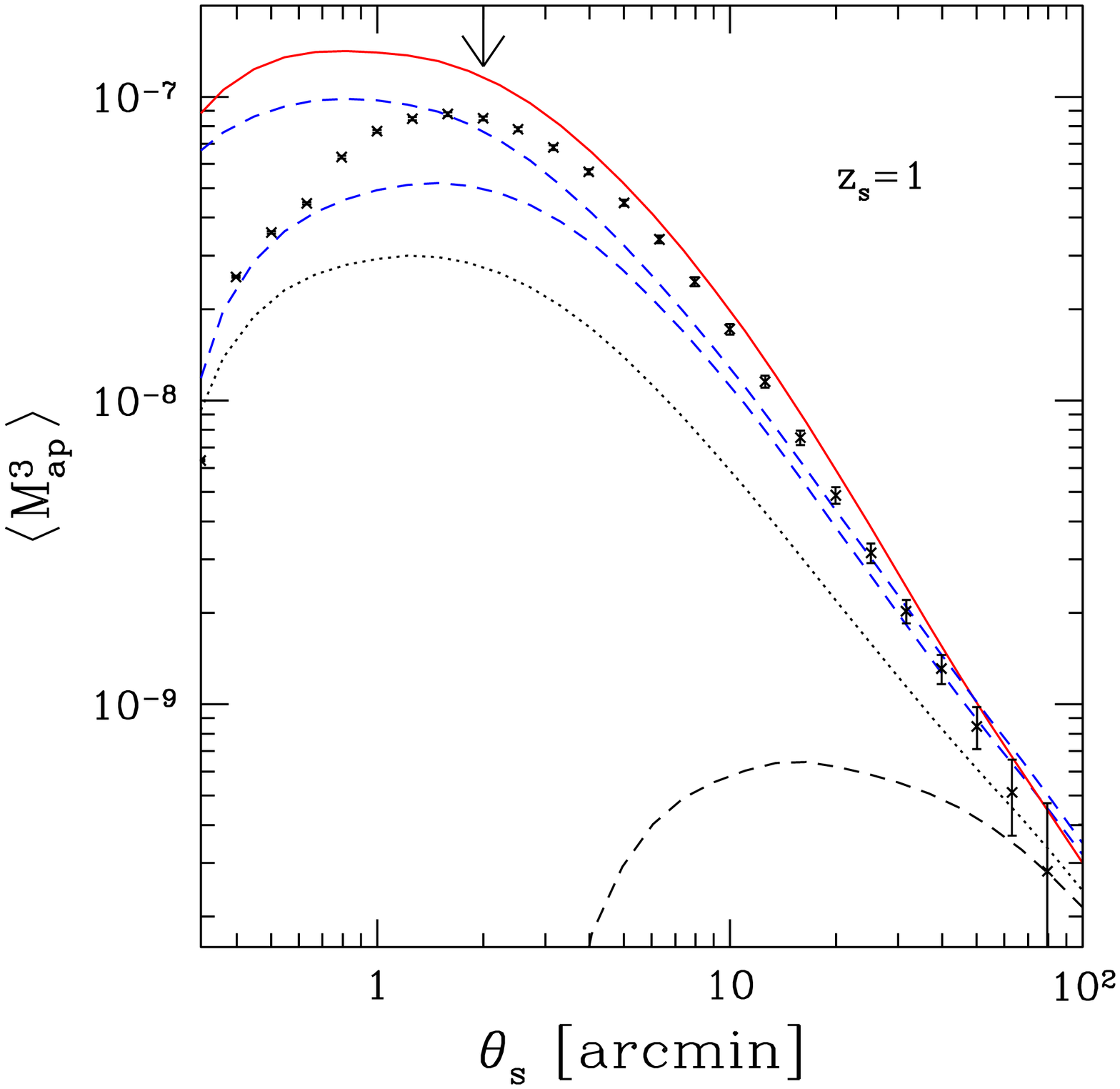}}
\epsfxsize=6.05 cm \epsfysize=5.4 cm {\epsfbox{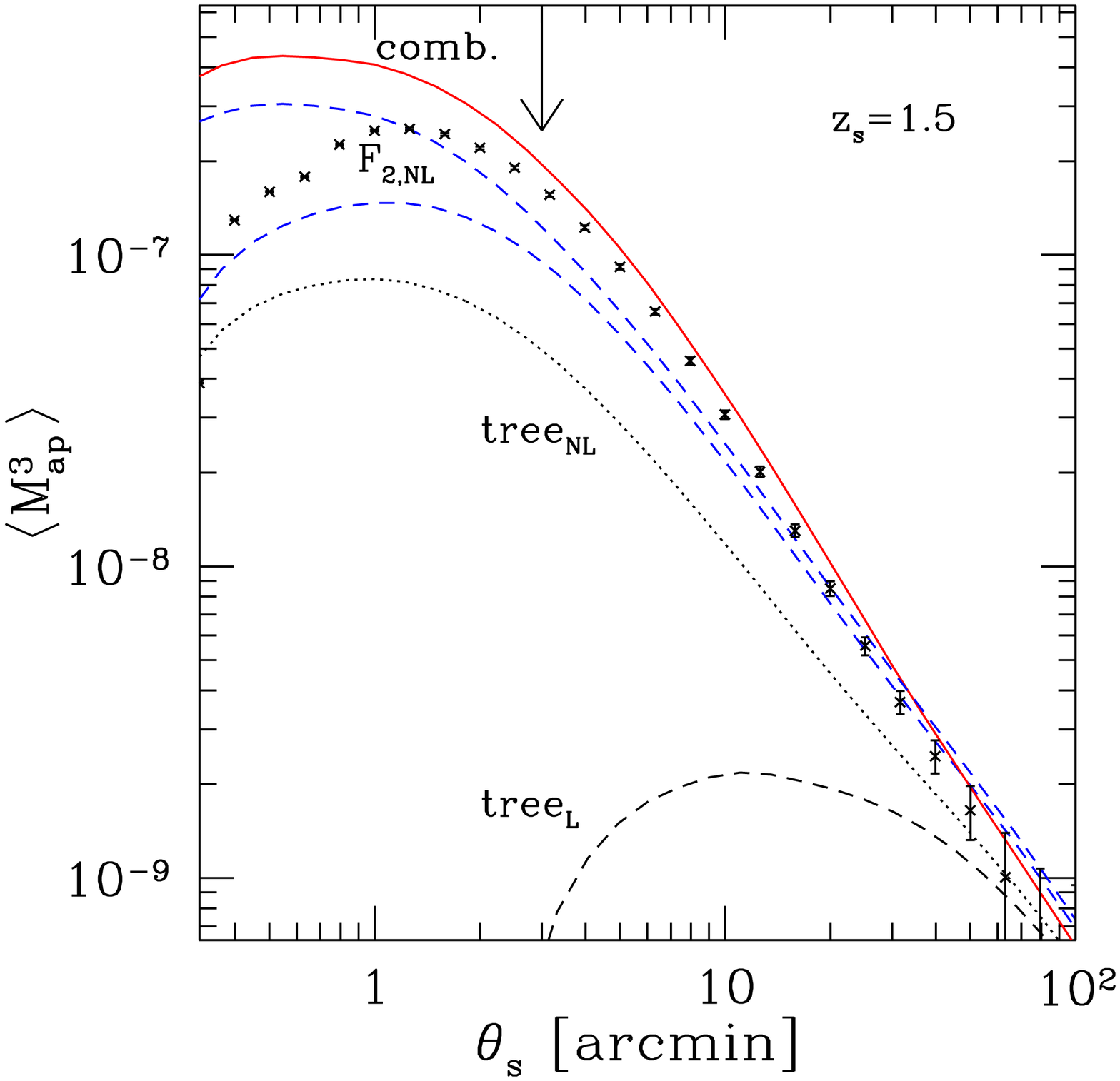}}
\end{center}
\caption{{\it Upper row:} convergence three-point correlation function
for equilateral triangles, as a function of the triangle side $\nu$, for sources
at redshifts  $z_s=0.6, 1$, and $1.5$.
The points are the results from numerical simulations with $3-\sigma$ error bars.
The low black dashed line ``$\rm tree_L$'', the black dotted line ``$\rm tree_{NL}$'' 
and the two blue dashed lines ``$F_{2,\rm NL}$'' are obtained from the ansatz
(\ref{B-treeL-def}), using either the linear 3D power spectrum, or the ``halo-fit'' power,
or an effective kernel $F_{2,\rm NL}$ with the ``halo-fit'' power (lower curve) or the 
power from our model (upper curve).
The red solid line ``comb.'' is our combined model, described in paper I.
The vertical arrows are at the same angular scale as in Fig.~\ref{fig_xi}.
{\it Middle row:} Third-order moment of the smoothed convergence, as a function
of the smoothing angle $\theta_s$.
{\it Lower row:} Third-order moment of the aperture mass, as a function
of the smoothing angle $\theta_s$.}
\label{fig_kappa3}
\end{figure*}

We now compare our analytical results with numerical simulations for three-point
functions. As in paper I, we also considered the predictions obtained from
the following three simple models, which have been used in some previous works.
(We did not consider the scale transformation introduced in \cite{Pan2007}
because we have already shown in paper I that it does not provide a sufficiently
accurate model for the convergence bispectrum. We have checked that we obtain
similar results for $\lag\kappa_s^3\rag$ and $\lag\Map^3\rag$.)

The first model, ``${\rm tree_L}$'', is the lowest-order (``tree-order'') prediction from
standard perturbation theory, which reads for the 3D bispectrum
\citep{Bernardeau2002}
\beq
B_{\rm tree_L}(k_1,k_2,k_3) \! = 2 F_2(k_1,k_2,\mu_{12}) \, P_L(k_1) P_L(k_2)
+ 2 \, {\rm cyc.} 
\label{B-treeL-def}
\eeq
where $\mu_{12}= (\vk_1\cdot\vk_2)/(k_1 k_2)$ and
\beq
F_2(k_1,k_2,\mu_{12}) = \frac{5}{7}
+ \frac{1}{2} \left(\frac{k_1}{k_2}\!+\!\frac{k_2}{k_1}\right) \mu_{12} 
+ \frac{2}{7} \, \mu_{12}^2 .
\label{F2-def}
\eeq
The second model, ``${\rm tree_{NL}}$'', is given by  Eq.(\ref{B-treeL-def}) where
we replace the linear 3D power $P_L(k)$ by the nonlinear power $P_S(k)$ from
\cite{Smith2003}.
The third model, ``$F_{2,\rm NL}$'', makes the additional modification to replace
the kernel $F_2$ by an effective kernel $F_{2,\rm NL}$ that interpolates from the
large-scale perturbative result (\ref{F2-def}) to a small-scale ansatz where
the angular dependence on $\mu_{12}$ vanishes, using the fitting formula from 
\cite{Scoccimarro2001a}.
We considered two variants, using either the ``halo-fit'' nonlinear power spectrum 
$P_S(k)$ from \cite{Smith2003} or the nonlinear power spectrum $P_{\rm tang}(k)$
of our model, see paper I and \cite{Valageas2011e}.

We show our results for the convergence three-point correlation function
$\zetakappa$ (for equilateral triangles) and for the third-order moments
$\lag\kappa_s^3\rag$ and $\lag\Map^3\rag$ in Fig.~\ref{fig_kappa3}.
In agreement with the second-order statistics shown in Fig.~\ref{fig_xi}, the
lowest-order perturbation theory prediction, ``${\rm tree_{L}}$'', remains valid down to
smaller angular scales for $\zetakappa$ and $\kappa_s$ than for $\Map$ because
the former involve uncompensated filters instead of a compensated filter.

The simulations slightly underestimate the power on moderate and large angular scales
because of the finite size of the simulation box ($240 h^{-1}$ and $480 h^{-1}$
Mpc, which corresponds to $343$ and $686$ arcmin at $z=1$), which cuts contributions from
longer wavelengths. This discrepancy
is not due to the analytic model because on the largest angular scales we can check that
all theoretical predictions converge on the linear theory (as they must for CDM power
spectra) while predicting somewhat more power than measured in the simulations
(see the first row in Fig.~\ref{fig_kappa3}). Therefore, this mismatch is not caused
by higher-order perturbative corrections (e.g., two-loop terms), which are even
smaller than the one-loop contributions that we included in our model. In agreement with this
explanation, the discrepancy is smaller for $\Map$ than for $\zetakappa$ and $\kappa$
because, for a given smoothing radius $\theta_s$, $\Map$ is less sensitive to larger
scales thanks to its compensated filter $\tW^{\Map}_{\theta_s}$.
This shows that analytical models, such as the one we propose here, are competitive
with numerical simulations if one needs to describe a broad range of scales.

In agreement with the results obtained in paper I for the convergence bispectrum,
using the nonlinear power within the ``tree-level'' expression (\ref{B-treeL-def}) significantly increases the third-order moments on smaller scales and improves
the general shape but is not sufficient to bridge the gap with the simulations.
Modifying the kernel $F_2$ by using the fitting formula of \cite{Scoccimarro2001a}
improves the predictions even more, especially when we use the 3D nonlinear power 
spectrum given by our model, which was shown earlier to be reasonably accurate
(see Fig.~\ref{fig_xi} and paper I and \cite{Valageas2011e}).
Indeed, as for the variance, the ``halo-fit'' power spectrum of \cite{Smith2003}
yields too little power on small scales, in agreement with previous works
\citep{Semboloni2011}. However, this approach still underestimates the
weak-lensing signal.

The best agreement with the numerical simulations is provided by our model.
As was seen for the convergence bispectrum in paper I, it is interesting to note
again the good match on the transition scales, $\theta_s \sim 5$ arcmin, which a priori are the most difficult
to reproduce since they are at the limit of validity of both perturbative approaches
(which break down at shell crossing) and halo models (which assume virialized halos).
On small angular scales we again predict more power than is measured in the
simulations, but like for the second-order moments this is at least partly caused by the lack
of small-scale power in the simulations because of the finite resolution.
Thus, we again plot in Fig.~\ref{fig_kappa3} the vertical arrows that were plotted
in Fig.~\ref{fig_xi}. Since the bispectrum typically scales as the square of the
power spectrum, this should roughly correspond to an accuracy threshold of
about $10\%$ for the simulations. We can check that our model agrees with the
numerical results down to this angular scale.
As in Fig.~\ref{fig_xi}, $\Map$ is much more sensitive than $\zetakappa$ and
$\kappa_s$ to this finite-resolution effect.

\section{Relative importance of the different contributions}
\label{contributions}

\begin{figure*}
\begin{center}
\epsfxsize=6.1 cm \epsfysize=5.4 cm {\epsfbox{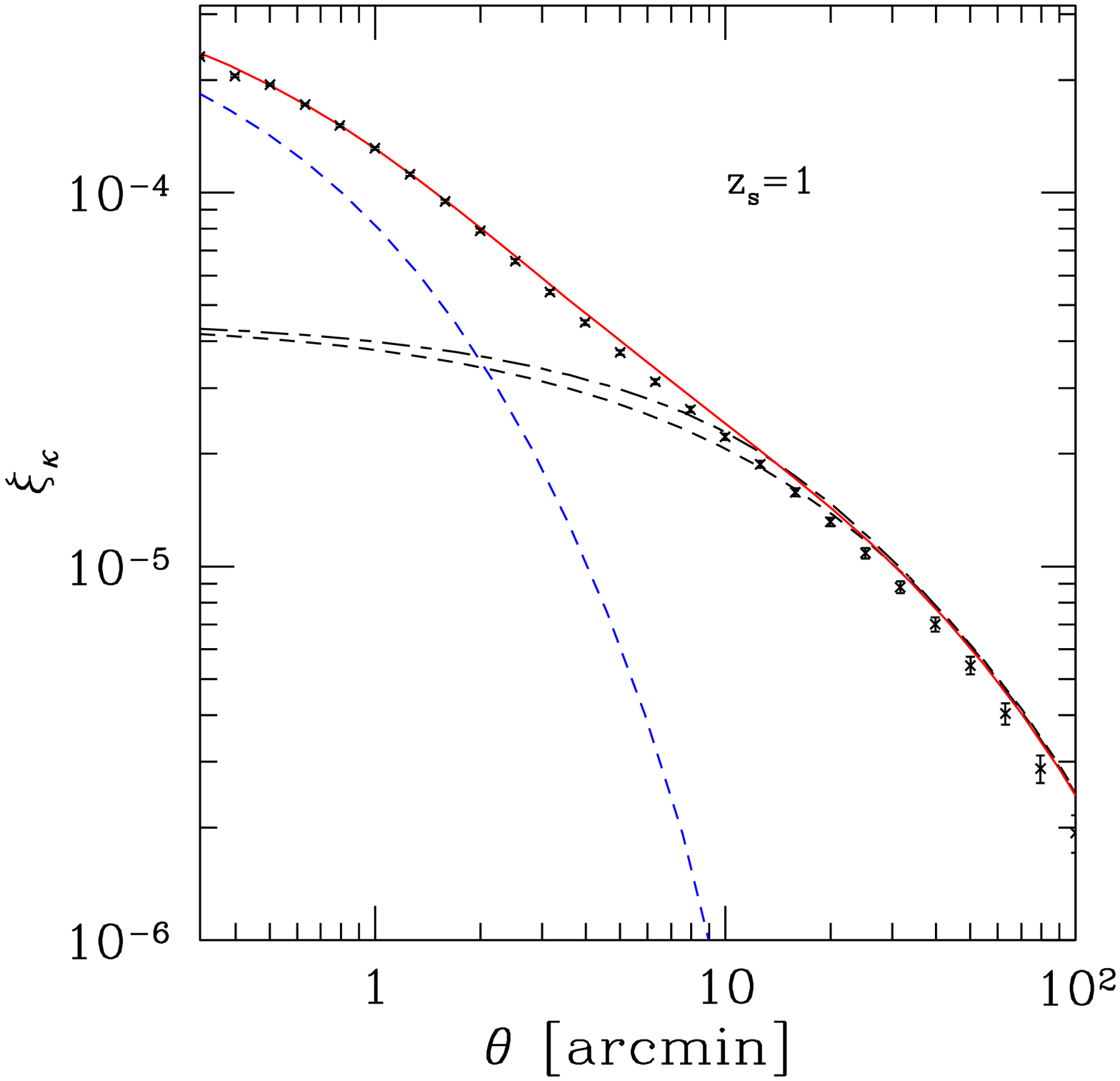}}
\epsfxsize=6.05 cm \epsfysize=5.4 cm {\epsfbox{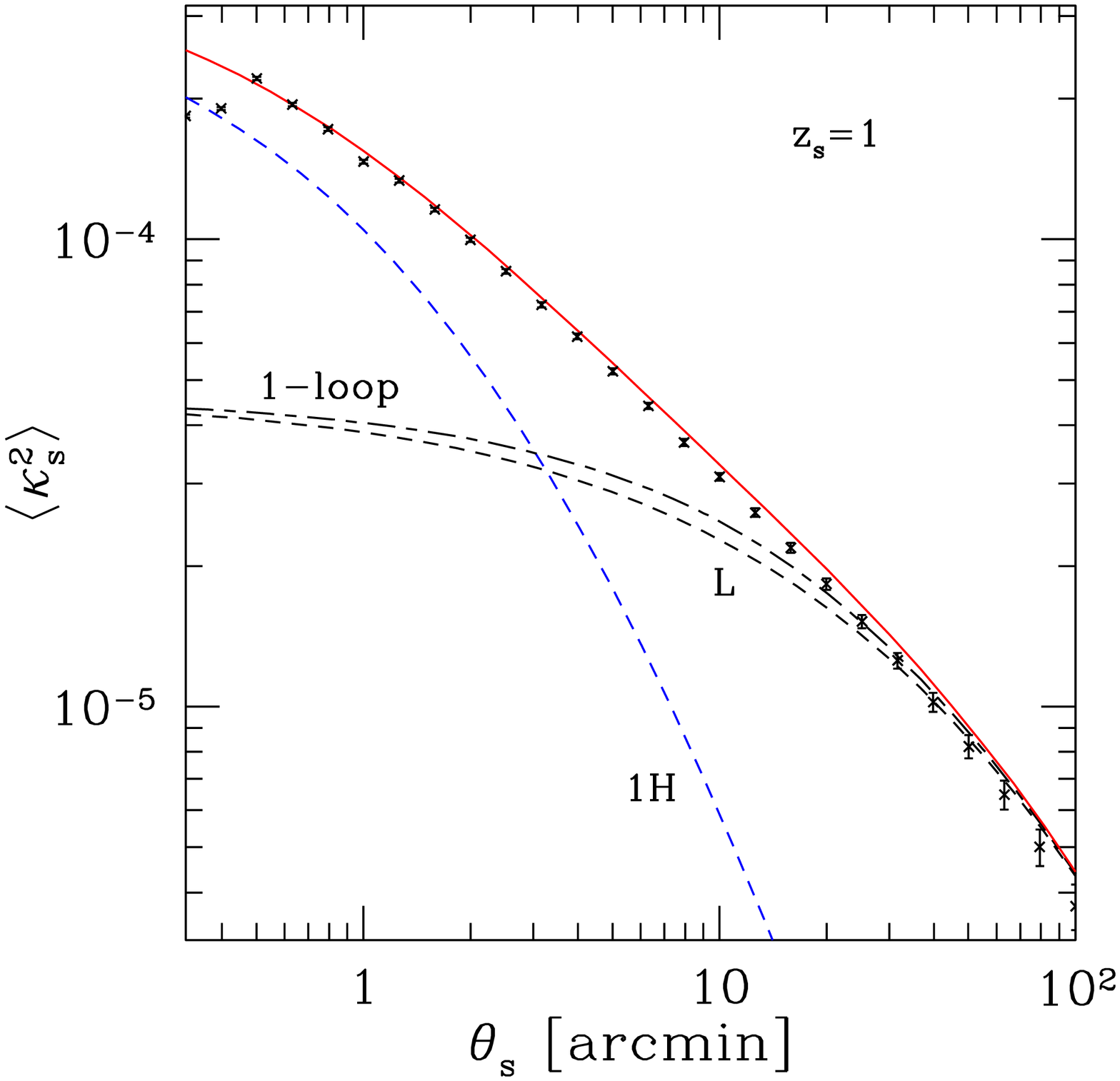}}
\epsfxsize=6.05 cm \epsfysize=5.4 cm {\epsfbox{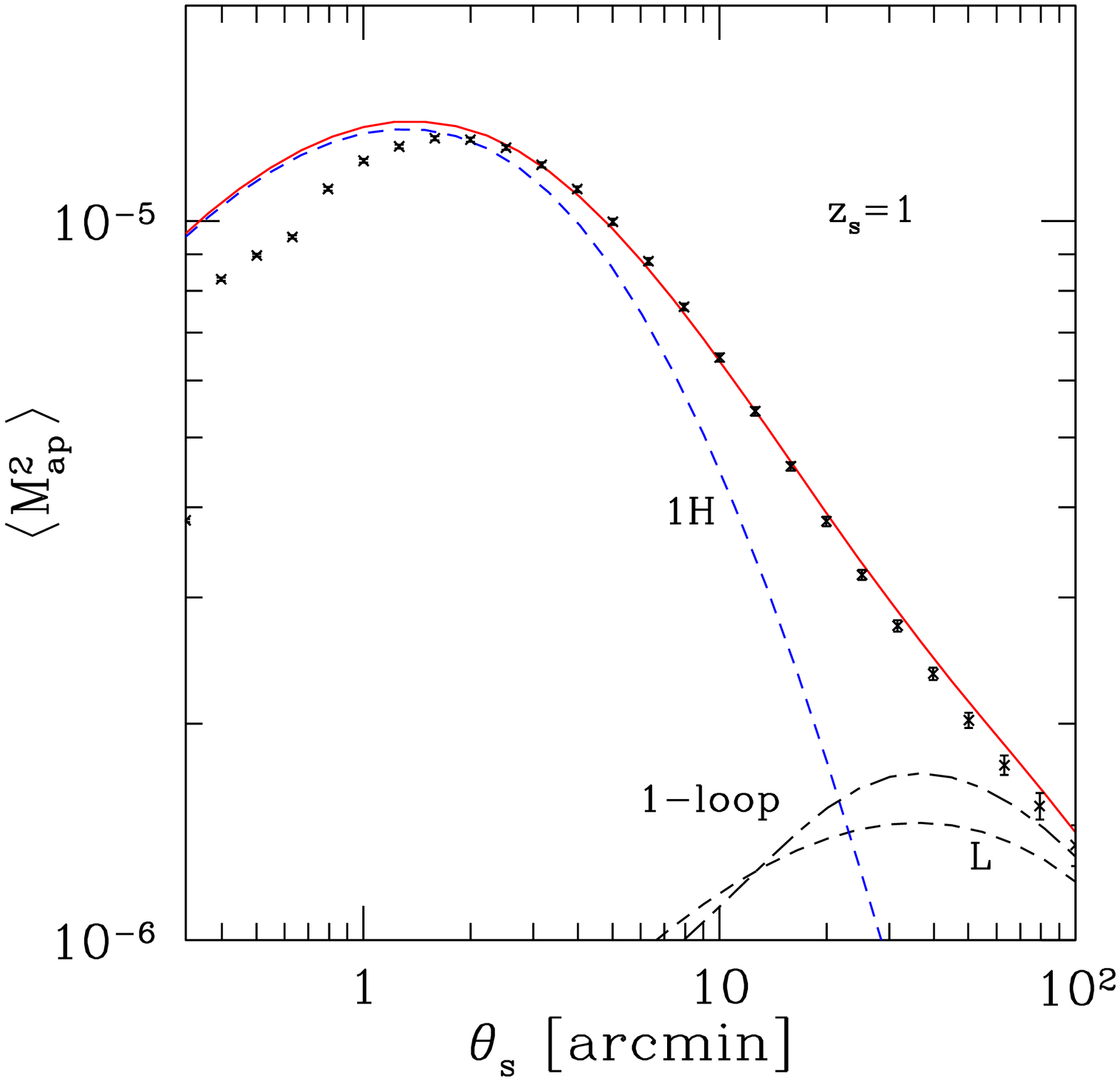}}
\end{center}
\caption{Convergence two-point correlation function (left panel), smoothed
convergence (middle panel), and aperture-mass (right panel) for sources at redshift
$z_s=1$. The points are the results from numerical simulations
with $3-\sigma$ error bars.
The low black dashed line ``L'' is the linear prediction, the middle black dash-dotted line
``1-loop'' is the two-halo contribution, for which we used a perturbative resummation that is 
complete up to one-loop order, the upper blue dashed line ``1H'' is the one-halo 
contribution, and the red solid line is our full model, as in Fig.~\ref{fig_xi}.}
\label{fig_xi_1H}
\end{figure*}

\begin{figure*}
\begin{center}
\epsfxsize=6.1 cm \epsfysize=5.4 cm {\epsfbox{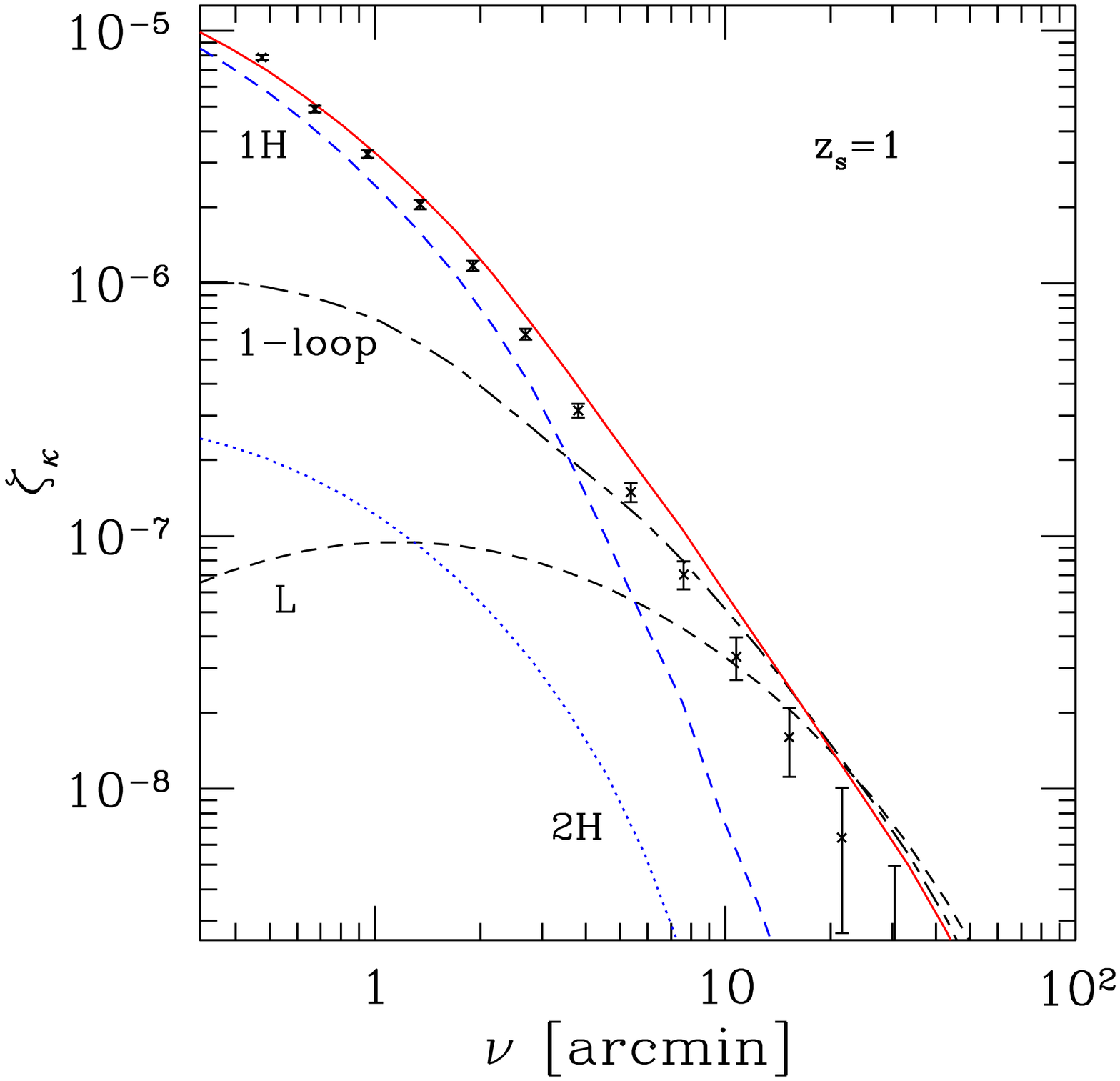}}
\epsfxsize=6.05 cm \epsfysize=5.4 cm {\epsfbox{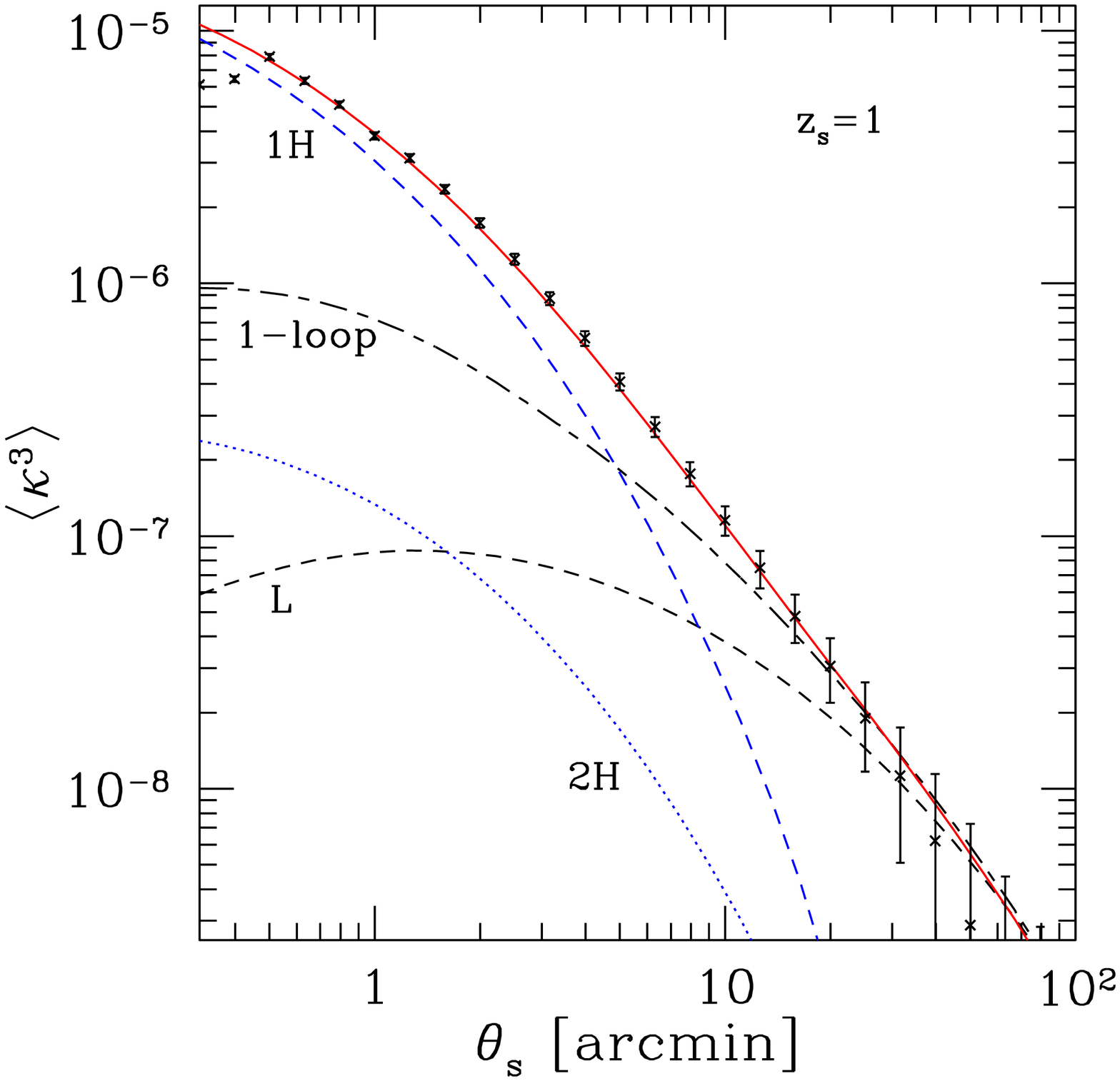}}
\epsfxsize=6.05 cm \epsfysize=5.4 cm {\epsfbox{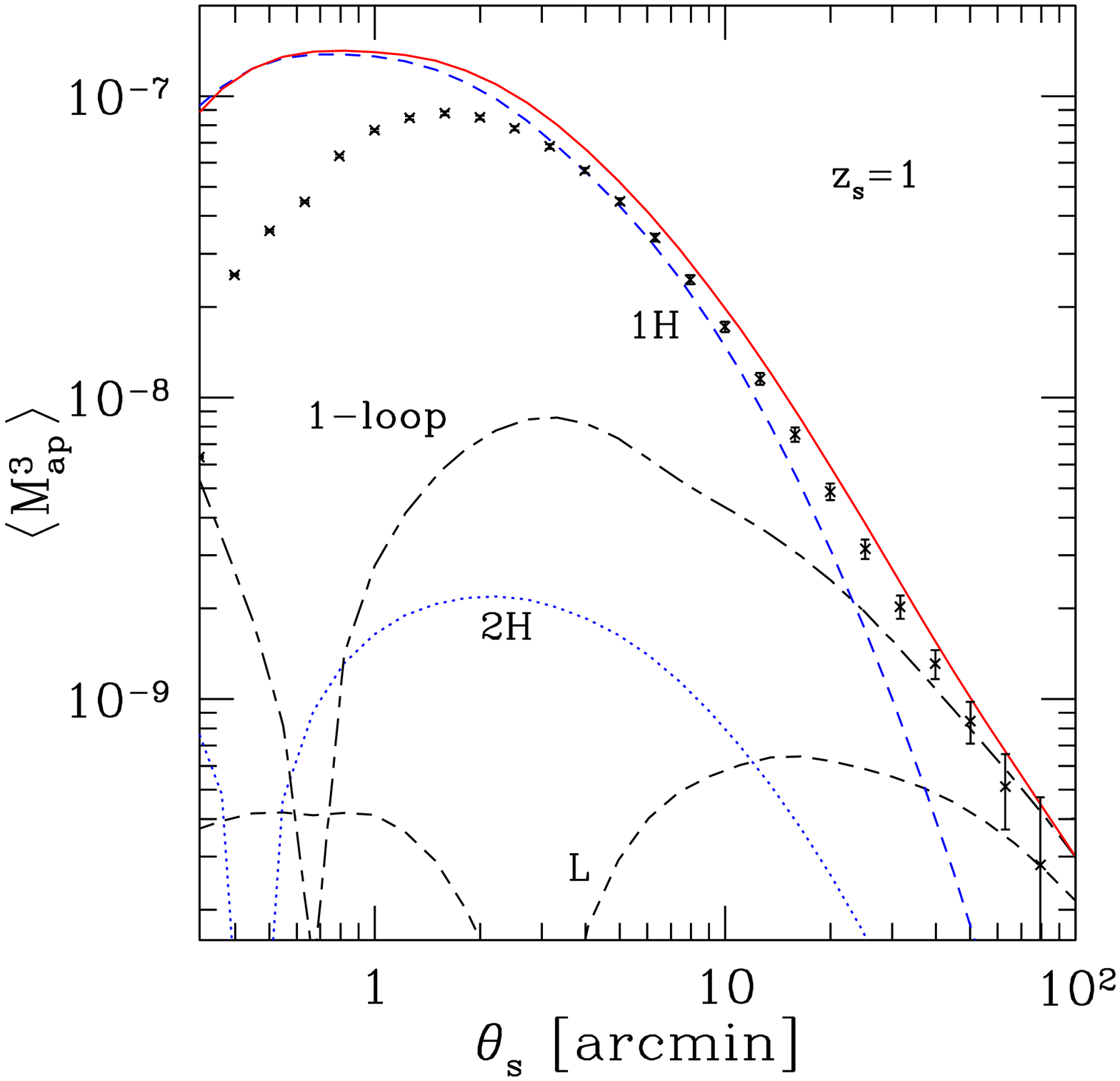}}
\end{center}
\caption{Convergence three-point correlation $\zetakappa$ for equilateral configurations
(left panel) and third-order moments $\lag\kappa_s^3\rag$ (middle panel) and
$\lag\Map^3\rag$ (right panel), for sources at redshift $z_s=1$. 
The points are the results from numerical simulations with $3-\sigma$ error bars.
The low black dashed line ``L'' is the lowest-order perturbative prediction (\ref{B-treeL-def}),
the ``1-loop'' dash-dotted black line is the prediction of one-loop standard perturbation
theory, which is identified with our three-halo term, the blue dotted line
``2H'' is the two-halo contribution and the upper blue dashed line
``1H'' is the one-halo contribution. The red solid line is our full model.}
\label{fig_kappa3_1H}
\end{figure*}

As in paper I, we have seen in the previous section that our model, which is
based on a
combination of perturbation theories and halo models, provides a good match to
numerical simulations. 
Therefore, it can be used to predict weak-lensing statistics for a variety of cosmologies, 
which is an important goal for observational and practical purposes.
A second use of our approach is to compare the different 
contributions that eventually add up to the signal that can be measured in
weak-lensing surveys. Thus, we can distinguish the various perturbative terms
as well as the nonperturbative contributions associated for instance with one-halo
or two-halo terms. This is a second advantage of these models, as compared with
fitting formulas or direct ray-tracing simulations.
This enables a deeper understanding of which properties of the matter distribution,
and of the overall cosmological setting, can be probed by a given gravitational
lensing measure. It can also help to estimate the accuracy that can be aimed at in
weak-lensing statistics as a function of scales, because different contributions suffer from 
different theoretical uncertainties.

\subsection{Two-point statistics}
\label{Two-point-1H}

We plot our results for $\xikappa$, $\lag\kappa_s^2\rag$,
and $\lag\Map^2\rag$ in Fig.~\ref{fig_xi_1H} at redshift $z_s=1$. 
In addition to the full model prediction that was already shown in Fig.~\ref{fig_xi}, we
show the underlying 2-halo and 1-halo contributions.
Because of their uncompensated filters, which make statistics at a given angular scale
receive contributions from 3D fluctuations on a wide range of scales, taking the
one-loop perturbative term into account only yields a small increase of $\xikappa$
and $\lag\kappa_s^2\rag$ over a wide range of angular scales, as compared with
the linear prediction.
In contrast, for $\lag\Map^2\rag$ this one-loop contribution peaks on a narrow range
of angular scales around $40$ arcmin and has a significant impact that improves
the match to the numerical results.
On small scales the two-halo contribution decreases close to the linear prediction thanks to
the partial resummation of higher perturbative orders. As explained in
\cite{Valageas2011d} and paper I, this is a useful improvement over the standard one-loop 
perturbation theory because it ensures that the two-halo term does not give significant contributions on very small scales, in agreement with physical expectations.

Then, these two-point weak-lensing quantities become dominated on small scales
by the
one-halo term but like for the 3D and 2D power spectra, there remains a significant
intermediate range.
Again, our model provides a satisfactory interpolation on these scales, but it would
be interesting to build a more systematic procedure, for instance by 
including higher orders of perturbation theory or by building a more refined matching 
between the two-halo and one-halo regimes.
In any case, Fig.~\ref{fig_xi_1H} clearly shows how $\xikappa$, $\lag\kappa_s^2\rag$,
and $\lag\Map^2\rag$ depend on large-scale perturbative density fluctuations or on
small-scale halo properties, as the angular scale varies.

We checked that we obtain similar results at redshifts $z_s=0.6$ and $1.5$.

\subsection{Three-point statistics}
\label{three-point-1H}

\begin{figure*}
\begin{center}
\epsfxsize=6.1 cm \epsfysize=5.4 cm {\epsfbox{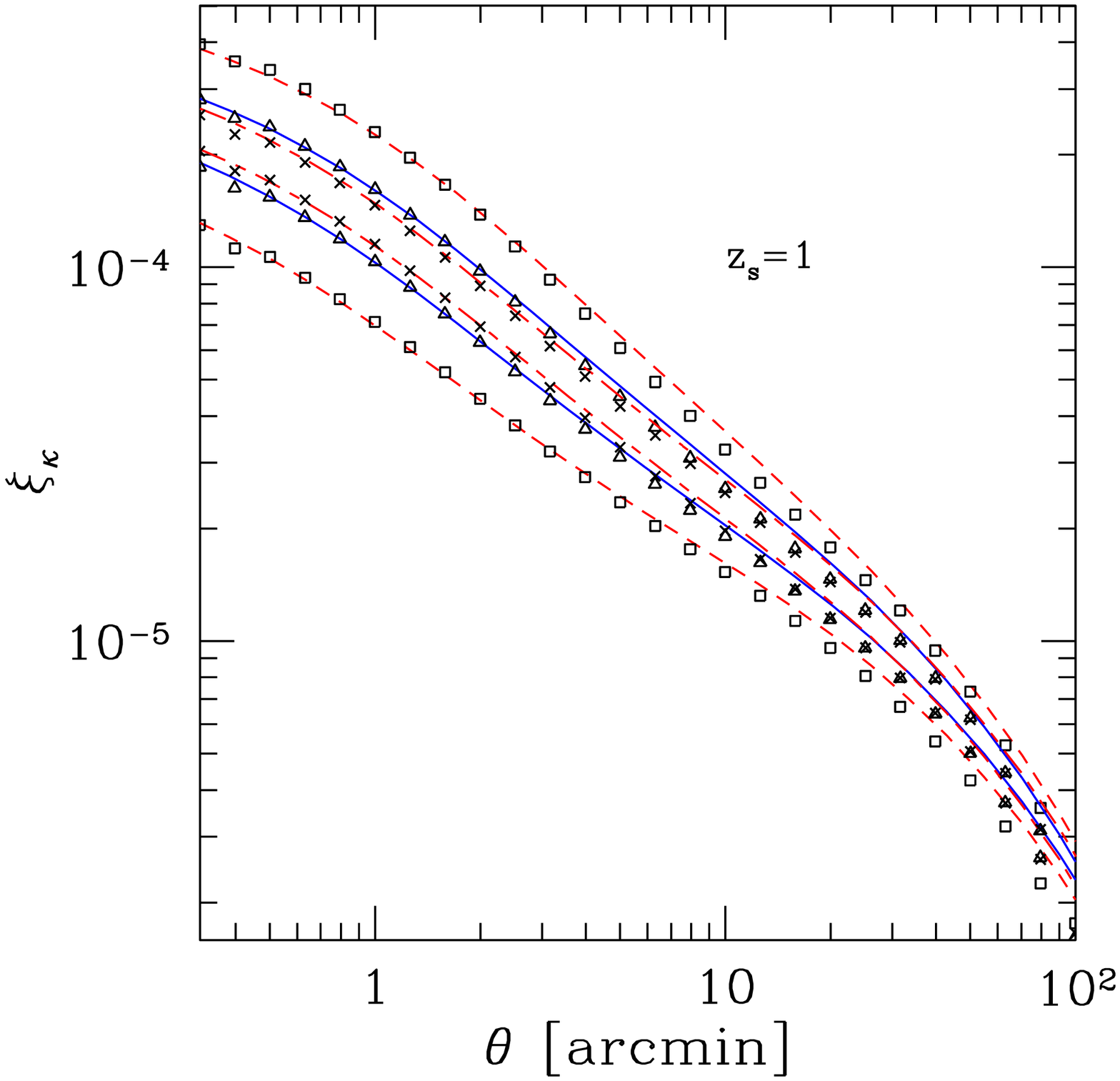}}
\epsfxsize=6.05 cm \epsfysize=5.4 cm {\epsfbox{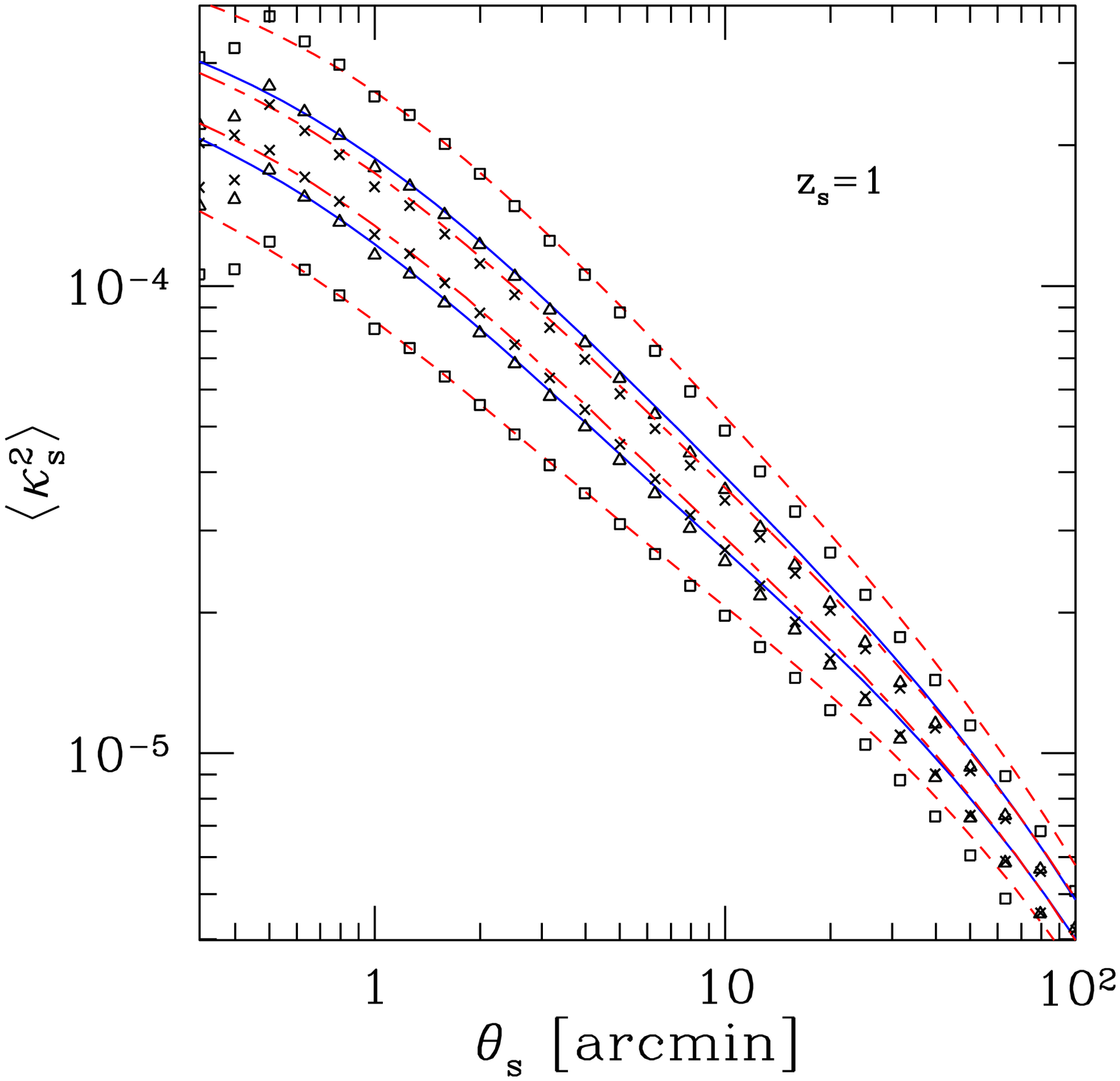}}
\epsfxsize=6.05 cm \epsfysize=5.4 cm {\epsfbox{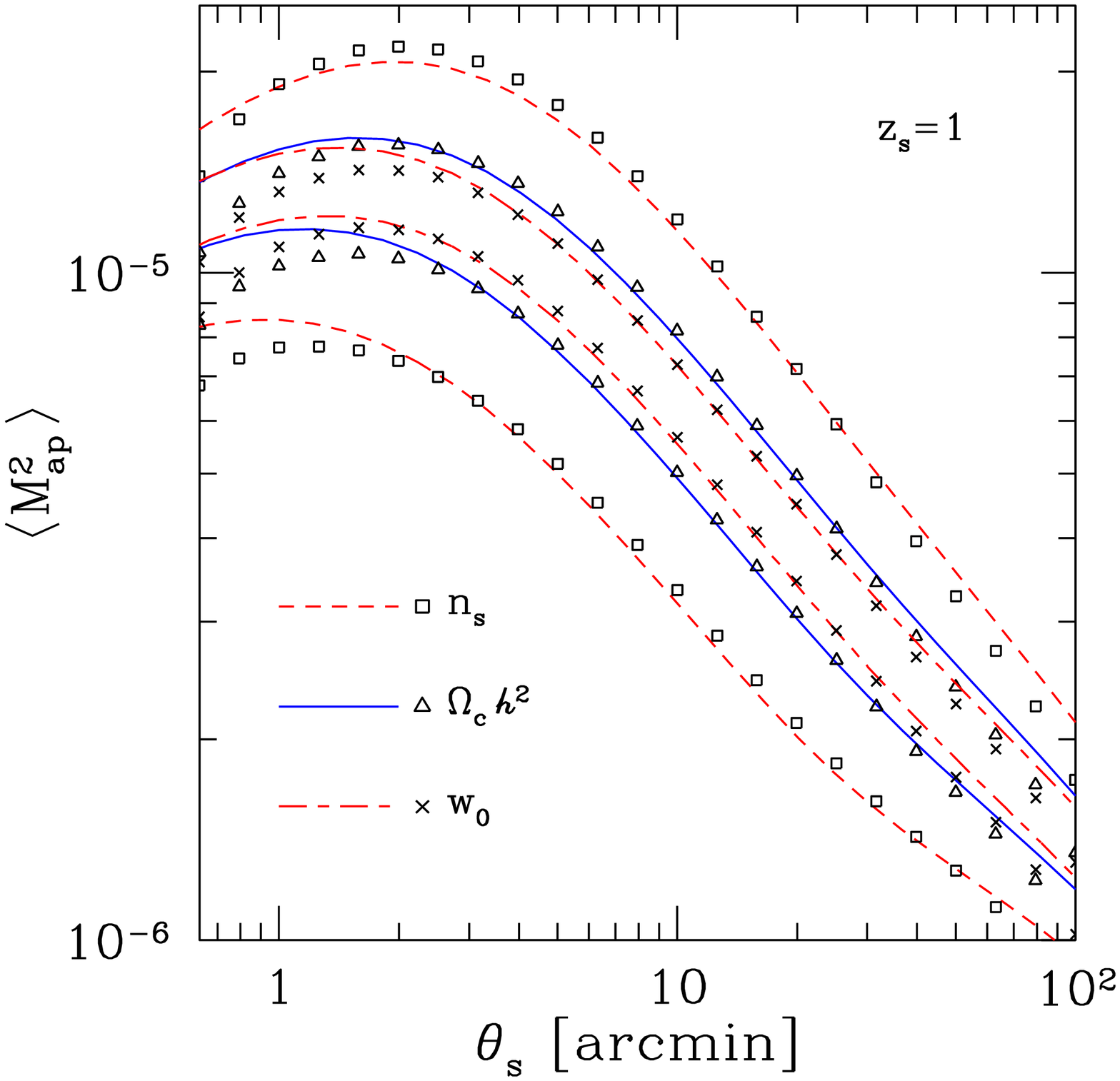}}
\end{center}
\caption{Convergence two-point correlation function (left panel), smoothed
convergence (middle panel), and aperture-mass (right panel) for sources at redshift
$z_s=1$ for six cosmologies. The points are the results from numerical simulations
and the lines are the predictions of our model.}
\label{fig_xi_comp}
\end{figure*}

\begin{figure*}
\begin{center}
\epsfxsize=6.1 cm \epsfysize=5.4 cm {\epsfbox{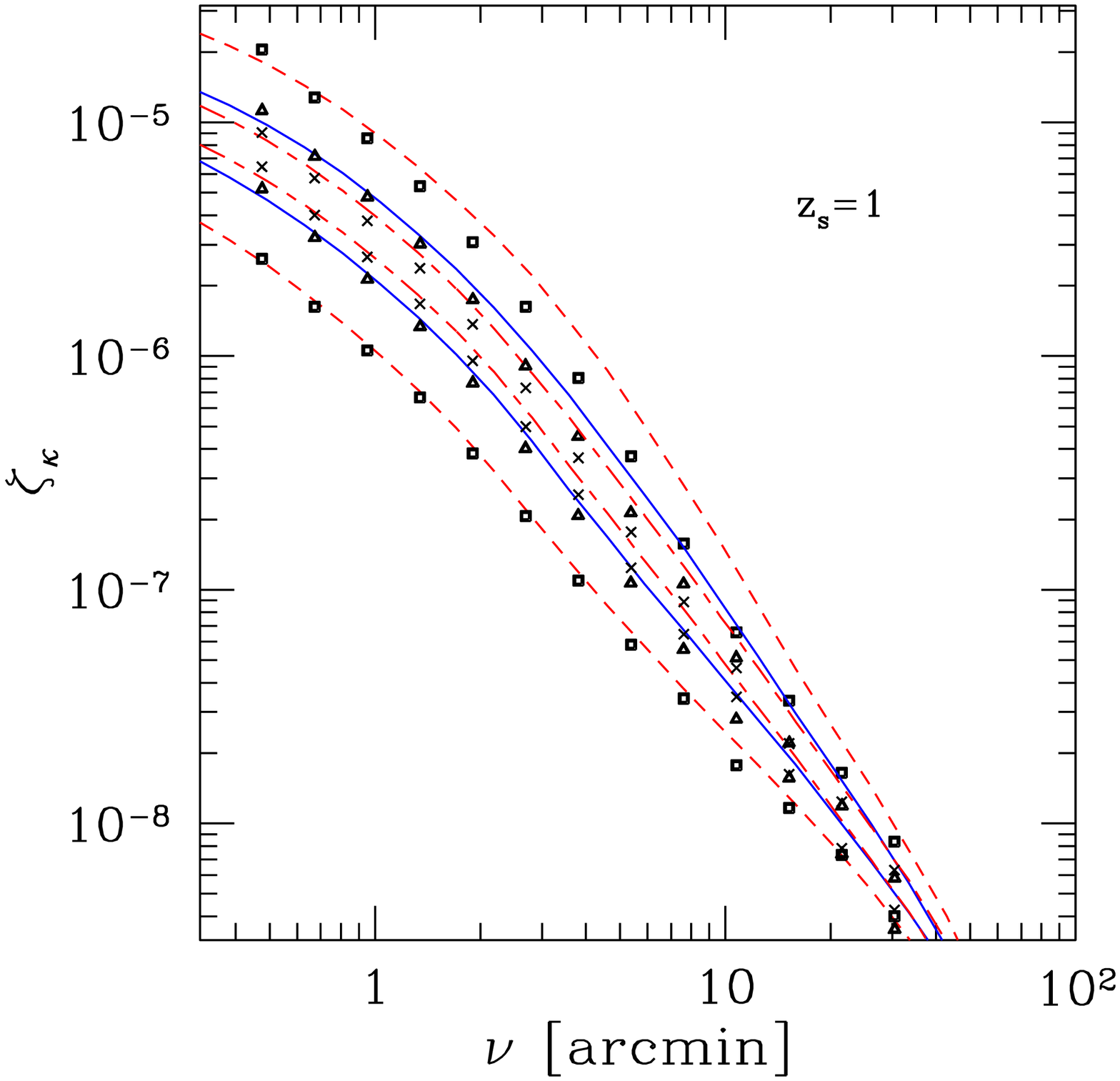}}
\epsfxsize=6.05 cm \epsfysize=5.4 cm {\epsfbox{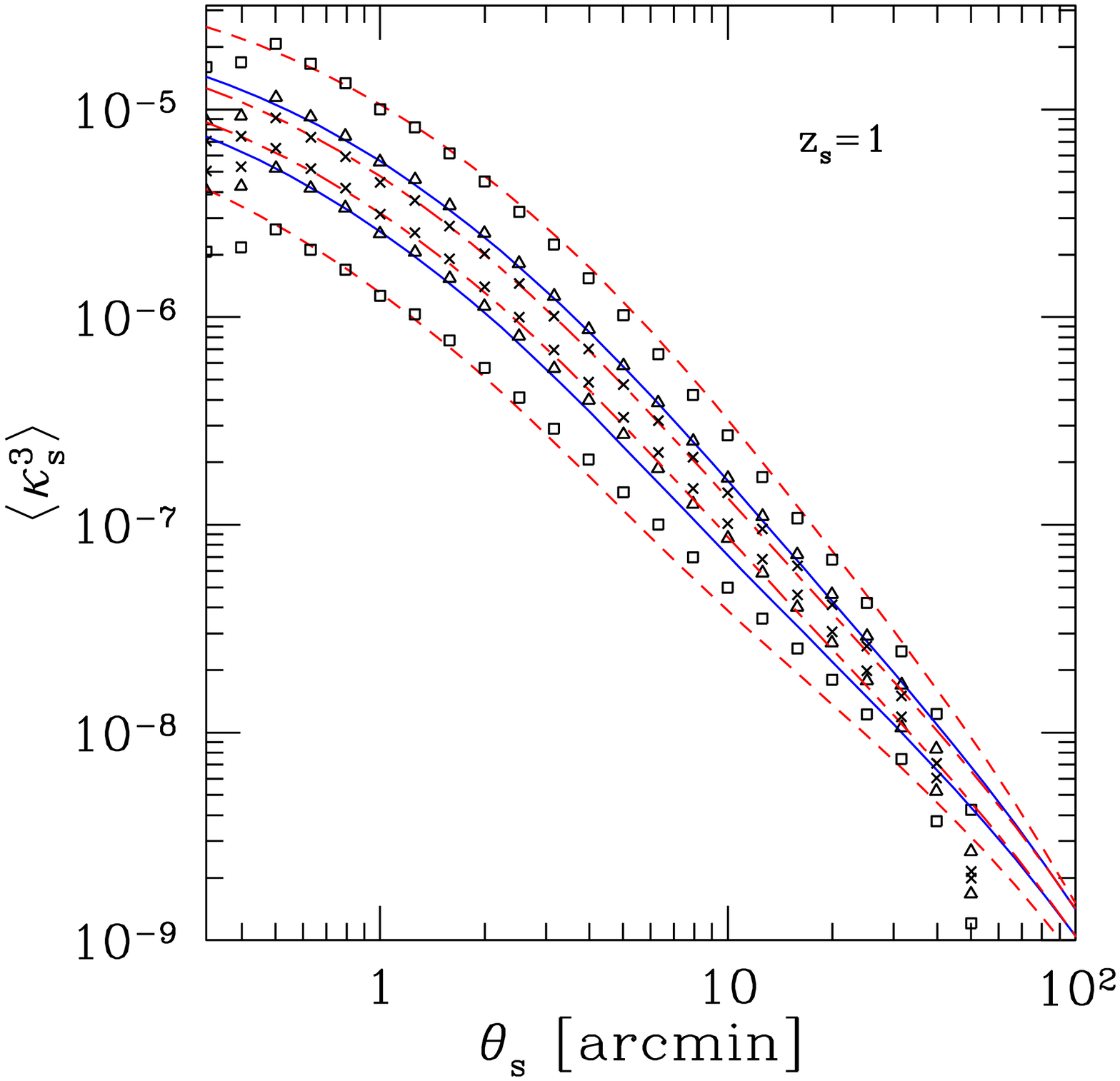}}
\epsfxsize=6.05 cm \epsfysize=5.4 cm {\epsfbox{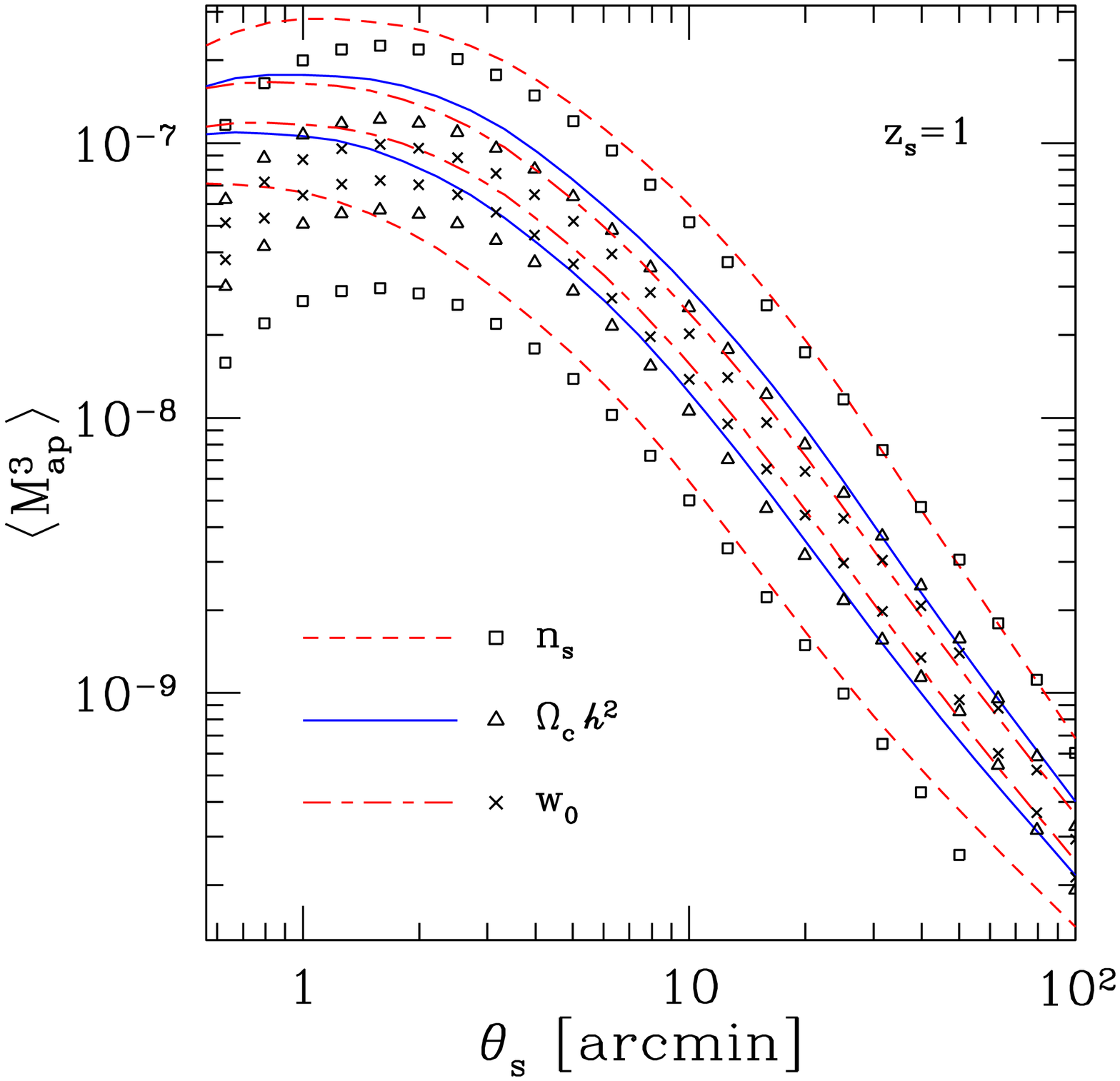}}
\end{center}
\caption{Convergence three-point correlation $\zetakappa$ for equilateral configurations
(left panel) and third-order moments $\lag\kappa_s^3\rag$ (middle panel) and
$\lag\Map^3\rag$ (right panel) for sources at redshift $z_s=1$ for six cosmologies.
The points are the results from numerical simulations and the lines are the predictions
of our model.}
\label{fig_kappa3_comp}
\end{figure*}

We now study the various contributions to the lensing three-point functions,
associated with the thee-halo, two-halo, and one-halo terms.

We plot our results for $\zetakappa$ (for equilateral configurations),
$\lag\kappa_s^3\rag$, and $\lag\Map^3\rag$ in Fig.~\ref{fig_kappa3_1H}, at redshift 
$z_s=1$. 
As in paper I, the three-halo term is identified with the perturbative
prediction and we used the standard perturbation theory at one-loop order. 
Like for the 3D bispectrum \citep{Valageas2011e}, and contrary
to the power spectrum, this gives a contribution that becomes
negligible on small scales, so that it is not necessary to use a resummation scheme
or to add a non-perturbative cutoff to ensure a good small-scale behavior.
On the other hand, contrary to the two-point statistics shown in Fig.~\ref{fig_xi_1H},
going to one-loop order now provides a great improvement over the tree-order result,
even for $\zetakappa$ and $\lag\kappa_s^3\rag$.
This feature was already noticed for the 3D bispectrum 
\citep{Sefusatti2010,Valageas2011e} and the convergence bispectrum
(paper I).
Thus, combining this one-loop perturbative contribution with the two-halo and
one-halo
terms is sufficient to obtain a good match to the simulations, from the quasilinear
to the highly nonlinear scales. This suggests that higher orders of perturbation theory
do not significantly contribute to the bispectrum and that we already have
a reasonably successful model. 
The two-halo term is also subdominant on all scales (by a factor $\sim 10$ at least).
This is a nice property since because it is a mixed term, which involves both large-scale
halo correlations and internal halo structures, it may be more difficult to predict
than the three-halo term (which is derived from systematic perturbation theories) and the one-halo
term (which only depends on internal halo profiles and mass function).
These various features were also observed for the 3D bispectrum \citep{Valageas2011e} and the convergence bispectrum (paper I).

Again, we checked that we obtain similar results at redshifts $z_s=0.6$ and $1.5$.

\section{Dependence on cosmology}
\label{Cosmology}

In this section we check the robustness of our model when we vary the cosmological 
parameters. As in paper I, we considered six alternative cosmologies, where 
$n_s$, $\Omega_{\rm c}h^2$, and $w_0$ are modified by $\pm 10\%$
with respect to the fiducial cosmology used in the previous sections.
The values of the associated cosmological parameters are given in
Table~I in App.~A of paper I.
We compare the predictions of our model with numerical simulations for these
six alternative cosmologies in Figs.~\ref{fig_xi_comp} and \ref{fig_kappa3_comp}
for the two-point and three-point statistics at $z_s=1$.
To avoid overcrowding the figures we did not plot the error bars of the numerical
simulations.
Each pair $n_s$, $\Omega_{\rm c}h^2$, and $w_0$ gives two curves that are roughly 
symmetric around the fiducial cosmology result, because we consider small deviations of $\pm 10\%$.
The deviations are largest for the $n_s$ case, which changes the
shape of the initial power spectrum as well as the normalization $\sigma_8$.
These six cases roughly cover the range that is allowed by current data,
and the $n_s$ case is already somewhat beyond the observational bounds
\citep{Komatsu2011}. Therefore, they provide a good check of the robustness of
our model for realistic scenarios.

Like for the Fourier-space statistics studied in paper I, the dependence on cosmology
of the two-point statistics is well reproduced by our model.
For the three-point statistics it is not easy to make a very precise comparison because
the numerical results show a greater level of noise and are sensitive to finite resolution
and finite size effects.
However, where the simulations are reliable, we also obtain a good match with our predictions.
We obtained similar results for $z_s=0.6$ and $z_s=1.5$, as well as for other
cosmologies where we vary $A_s$ or $\Omega_{\rm de}$ by $\pm 10\%$.
This shows that our model and, more generally, models based on combinations
of perturbation theory and halo models provide a good modeling of the matter
distribution and of weak gravitational lensing effects and capture their dependence
on cosmology.
Moreover, Figs.~\ref{fig_xi_comp} and \ref{fig_kappa3_comp} clearly show that
this analytical modeling is competitive with current ray-tracing simulations,
because it provides reliable predictions over a greater range of scales.
In particular,  Figs.~\ref{fig_xi_comp} and \ref{fig_kappa3_comp} show that
the accuracy of our model is sufficient to constrain $n_s$, $\Omega_ch^2$,
and $w_0$ to better than $10\%$.

\section{Multi-scale moments}
\label{multi-scale}

\begin{figure}
\begin{center}
\epsfxsize=8 cm \epsfysize=5.6 cm {\epsfbox{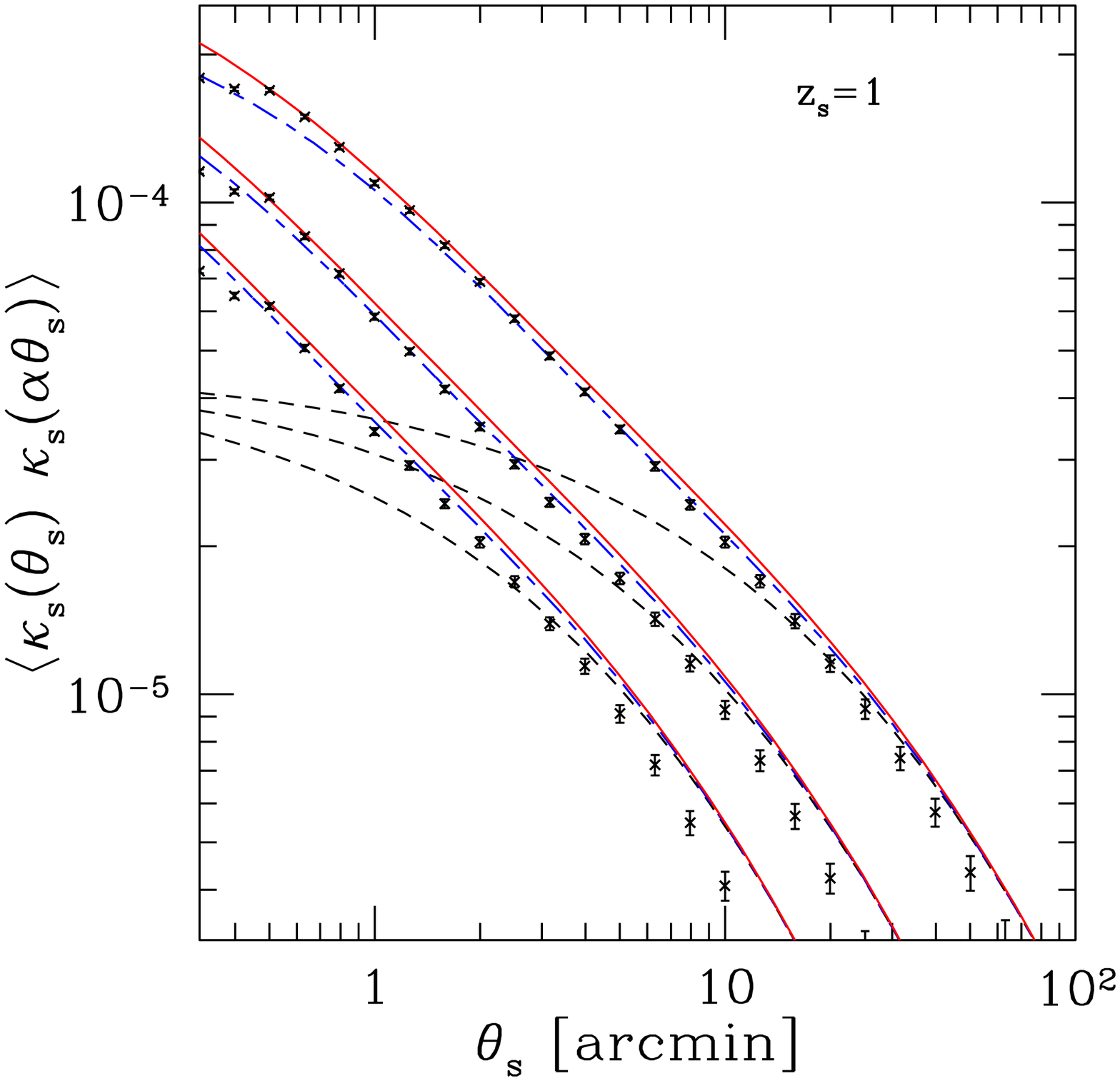}} \\
\epsfxsize=8 cm \epsfysize=5.6 cm {\epsfbox{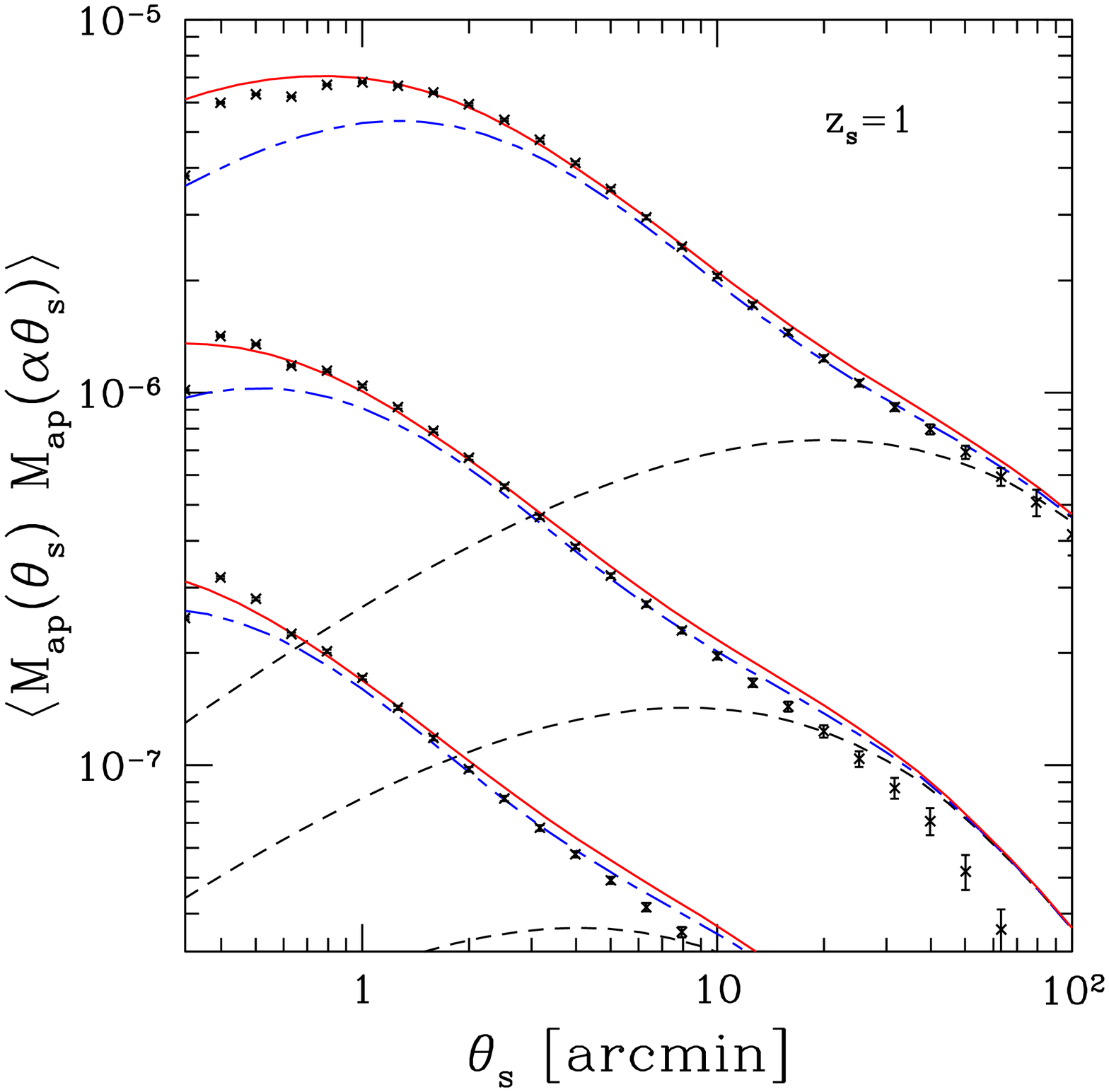}}
\end{center}
\caption{Two-scale second-order moments
$\lag \kappa_s(\theta_s)\kappa_s(\alpha\theta_s)\rag$ (upper panel) and
$\lag \Map(\theta_s)\Map(\alpha\theta_s)\rag$ (lower panel)
as a function of $\theta_s$, at $z_s=1$. We show the cases $\alpha=2, 5$, and $10$
from top to bottom.
The symbols are the same as in Fig.~\ref{fig_xi}.}
\label{fig_alpha_2}
\end{figure}

\begin{figure}
\begin{center}
\epsfxsize=8 cm \epsfysize=5.6 cm {\epsfbox{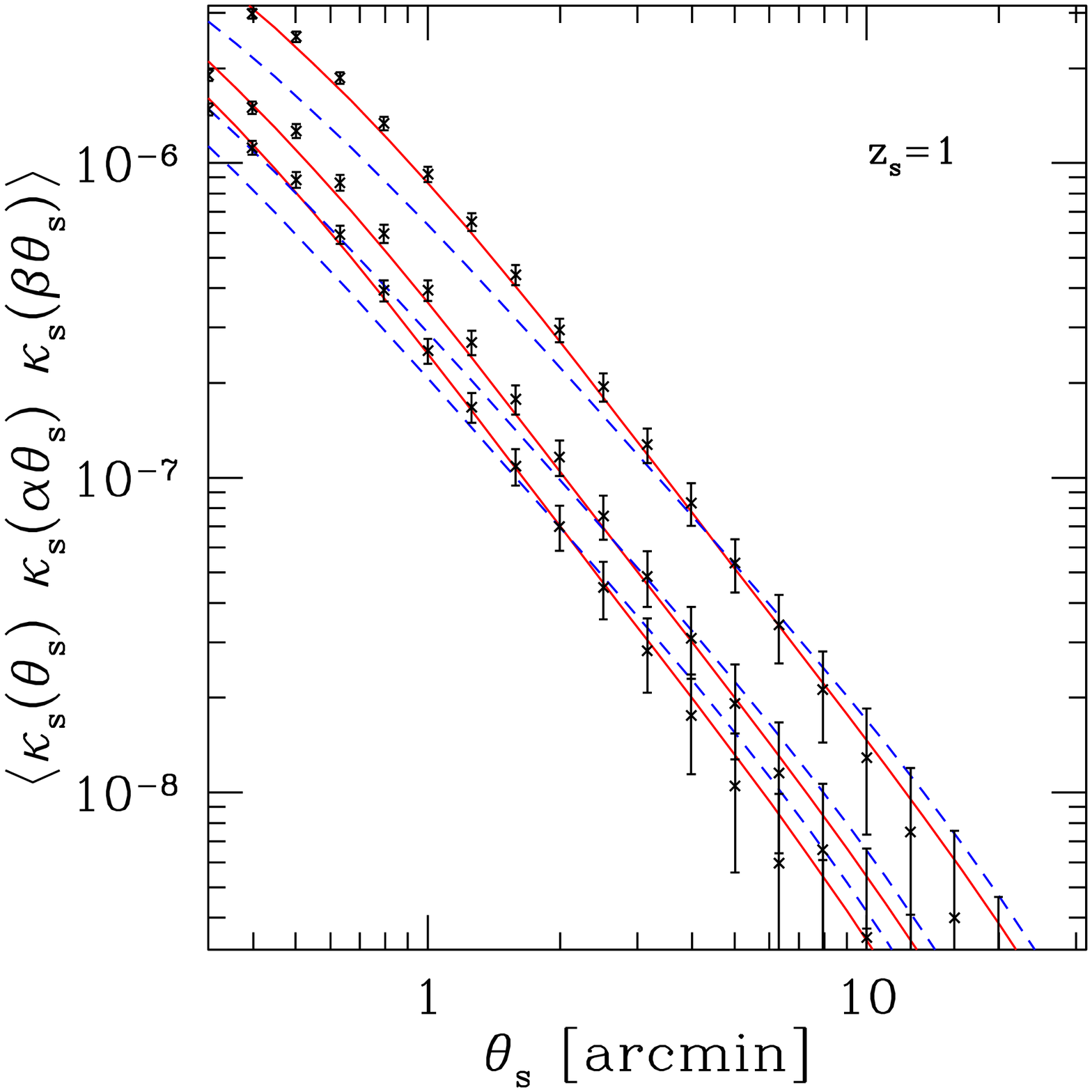}} \\
\epsfxsize=8 cm \epsfysize=5.6 cm {\epsfbox{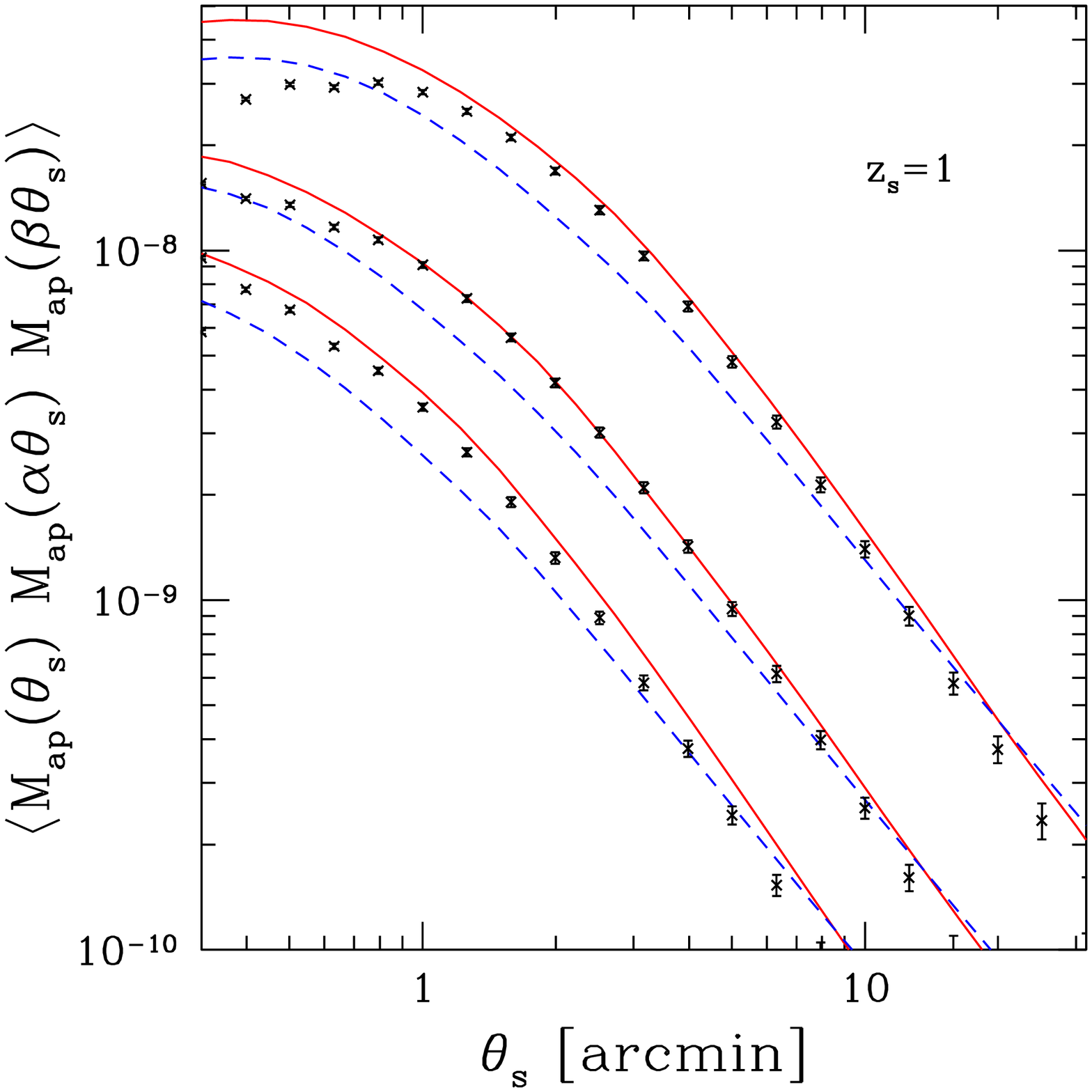}}
\end{center}
\caption{Three-scale third-order moments at $z_s=1$ as a function of $\theta_s$.
{\it Upper panel:} 
$\lag \kappa_s(\theta_s)\kappa_s(\alpha\theta_s)\kappa_s(\beta\theta_s)\rag$.
{\it Lower panel:} 
$\lag \Map(\theta_s)\Map(\alpha\theta_s)\Map(\beta\theta_s)\rag$.
We show the cases
$\{\alpha,\beta\}=\{2,5\}$, $\{3,9\}$, and $\{5,10\}$ from top to bottom.
The blue dashed line is the  ``$F_{2,\rm NL}$'' ansatz (\ref{B-treeL-def}), using the
effective kernel $F_{2,\rm NL}$ from \cite{Scoccimarro2001a} and the power from
our model (as in the upper blue dashed line in Fig.~\ref{fig_kappa3}), while the red 
solid line is our model.}
\label{fig_alphabeta_3}
\end{figure}

In the previous sections we considered single-scale moments,
$\lag X_s(\theta_s)^p\rag$, associated with one smoothing
window $W_{\theta_s}^{X_s}$ with a single angular radius
$\theta_s$. One can also use multi-point statistics such as
$\lag X_s(\vtheta_1;\theta_{s1}) .. X_s(\vtheta_p;\theta_{sp})\rag$ 
associated with $p$ windows centered on $p$ different directions
$\vtheta_i$ and with $p$ different radii $\theta_{si}$.
In this section, we briefly check the validity of our model for
the case of centered multi-scale moments. To do this, all quantities
$X_{si}$ are centered on the same direction on the sky but we
allow the angular radii $\theta_{si}$ to be different.
Then, Eqs.(\ref{Xs-variance}) and (\ref{Xs-3-real}) generalize
as
\beqa
\lag X_s(\theta_{s1}) X_s(\theta_{s2})\rag & \!\!= \!\! & 2\pi \!
\int_0^{\infty} \!\! \dd\ell \, \ell \, \Pkappa(\ell) \,
\tW^{X_s}_{\theta_{s1}}(\ell \theta_{s1}) 
\tW^{X_s}_{\theta_{s2}}(\ell \theta_{s2}) \nonumber \\ 
&&
\label{Xs-cross-2}
\eeqa
and
\beqa
\lag X_s(\theta_{s1}) X_s(\theta_{s2}) X_s(\theta_{s3})\rag & = & 
4\pi \int_0^{\theta_{s1}} \frac{\dd\theta_1 \theta_1}{\pi\theta_{s1}^2}
W^{X_s}_{\theta_{s1}}(\theta_1) \nonumber \\ 
&& \hspace{-3cm} \times \int_0^{\theta_{s2}} \!\!
\frac{\dd\theta_2 \theta_2}{\pi\theta_{s2}^2} W^{X_s}_{\theta_{s2}}(\theta_2) 
\int_0^{\theta_{s3}} \frac{\dd\theta_3 \theta_3}{\pi\theta_{s3}^2}
W^{X_s}_{\theta_{s3}}(\theta_3) \nonumber \\ 
&& \hspace{-3cm} \times \int_0^{\pi} \!\!\! \dd\varphi_2
\int_0^{2\pi} \!\!\! \dd \varphi_3 \, \zetakappa(\nu_1,\nu_2,\nu_3) .
\eeqa

We show in Fig.~\ref{fig_alpha_2} the two-scale second-order moments
$\lag \kappa_s(\theta_s)\kappa_s(\alpha\theta_s)\rag$ and
$\lag \Map(\theta_s)\Map(\alpha\theta_s)\rag$ for a scale-ratio
$\alpha=2, 5$, and $10$, at $z_s=1$. A higher $\alpha$ yields a
smaller moment because it corresponds to a larger second angular radius
$\alpha\theta_s$. 
We obtain the same level of agreement as for the single-scale variances
shown in Fig.~\ref{fig_xi}. 
In particular, we obtain a good match on small angular scales where the
``halo-fit'' formula somewhat underestimates the weak-lensing power.
On large scales our analytical results are somewhat larger than the data obtained
from the numerical simulations. This is due to the missing of large-scale modes
in the simulations because of the finite size of the simulation boxes 
($240 h^{-1}$ and $480 h^{-1}$ Mpc) that correspond to $343$ and $686$ arcmin
at $z=1$). 
This is more apparent than in the single-scale plots of Fig.~\ref{fig_xi} because
we probe larger scales since the factor $\alpha$ is greater than unity.
This again shows the advantage of analytical models such as ours, which are
competitive with current ray-tracing numerical simulations to describe a broad
range of scales.

We show in Fig.~\ref{fig_alphabeta_3} the three-scale third-order moments
$\lag \kappa_s(\theta_s)\kappa_s(\alpha\theta_s)\kappa_s(\beta\theta_s)\rag$
and
$\lag \Map(\theta_s)\Map(\alpha\theta_s)\Map(\beta\theta_s)\rag$
for the scale-ratios $\{\alpha,\beta\}=\{2,5\}$, $\{3,9\}$, and $\{5,10\}$,
at $z_s=1$.
Higher values of $\alpha$ and $\beta$ give a smaller moment since they
correspond to larger second and third angular radii $\alpha\theta_s$
and $\beta\theta_s$. 
We obtain the same level of agreement as for the single-scale third-order
moments shown in Fig.~\ref{fig_kappa3}.
In particular, our model recovers the dependence on the ratios $\{\alpha,\beta\}$
and on the scale $\theta_s$ and performs better than the other models studied in
this paper.
To simplifiy the figures, we show in Fig.~\ref{fig_alphabeta_3} only the second-best
model, i.e. the ``$F_{2,\rm NL}$'' ansatz (\ref{B-treeL-def}) using the
effective kernel $F_{2,\rm NL}$ from \cite{Scoccimarro2001a} and the power from
our model (as in the upper blue dashed line in Fig.~\ref{fig_kappa3}).
Other models show similar behaviors to those found in Fig.~\ref{fig_kappa3}
for single-scale moments (i.e., a lack of power on moderate and small angular scales).
As for the second-order statistics, the underestimate of the weak-lensing signal 
by the simulations appears at a smaller angle $\theta_s$ than in the single-scale
case shown in Fig.~\ref{fig_kappa3} because the factors $\alpha$ and $\beta$
are larger than unity and increase the sensitivity to larger scales at fixed $\theta_s$.

\section{Conclusion}
\label{Conclusion}

We have investigated the performance of current theoretical modeling
of the 3D matter density distribution with respect to weak-lensing statistics,
focusing on configuration-space statistics, specifically the convergence
two-point and three-point correlation functions and the second- and third-order
moments of the smoothed convergence and of the aperture mass.
As in paper I, where we studied Fourier-space statistics, 
we found that a model introduced in previous works
\citep{Valageas2011d,Valageas2011e}, which combines the (resummed) one-loop perturbation theory
with a halo model, fares better than some other recipes based on fitting formulae
to numerical simulations or more phenomenological approaches.
It yields a reasonable agreement with numerical simulations and provides
a competitive approach, because it remains difficult and time-consuming to describe
a range of scales that spans three orders of magnitude or more by ray-tracing 
simulations.

One advantage of our approach compared with numerical simulations or fitting
formulas is that it allows us to decompose the integrated weak-lensing signal
over several contributions that are associated with specific properties of the underlying
3D density field. Thus, we can distinguish perturbative terms, which can be derived
from perturbation theory, from nonperturbative terms that are associated for instance
with one-halo contributions, which depend on the density profile and mass function
of virialized halos. This is useful because i) these different terms suffer from different
theoretical uncertainties and ii) it allows one to understand which aspects of the
matter distribution are probed by weak-lensing statistics, while angular scales vary.

Like for the Fourier-space statistics studied in paper I and the 3D statistics studied
in \citet{Valageas2011d,Valageas2011e}, we found that including one-loop terms in
the perturbative contribution brings a more significant improvement compared with
the lowest-order perturbation theory for three-point statistics than for two-point statistics.
Then, while large scales are described by these perturbative contributions and
small scales by one-halo contributions, the nonperturbative two-halo term that gives
an additional contribution to three-point statistics is always subdominant.
This is a nice property because this mixed term is more difficult to model and
may be less accurate than other contributions (see also paper I). 

Consequently, our model provides reliable predictions for weak-lensing statistics,
from small to large scales, and for a variety of cosmologies.
It could still be improved in various manners. First, the accuracy of the perturbative 
contribution may be increased by including higher orders beyond one-loop or
by using alternative resummation schemes. Second, the underlying halo model
could be refined to include substructures \citep{Sheth2003a,Giocoli2010},
deviations from spherical profiles \citep{Jing2002,Smith2006}, or the effect
of baryons \citep{Guillet2010}.
Next, the model could be generalized to non-Gaussian initial conditions, which 
yield distinctive signatures in the bispectrum \citep{Sefusatti2010}.

\acknowledgement

We would like to thank Takashi Hamana for helpful discussions.
M.S. and T.N. are supported by a Grant-in-Aid for the Japan Society for Promotion of
Science (JSPS) fellows. This work is supported in part by the French
``Programme National de Cosmologie et Galaxies'' and the
French-Japanese ``Programme Hubert Curien/Sakura, projet 25727TL'',
the JSPS Core-to-Core Program ``International Research Network for Dark Energy'', 
a Grant-in-Aid for Scientific Research on Priority Areas No. 467 ``Probing the Dark
Energy through an Extremely Wide and Deep Survey with Subaru
Telescope'', a Grant-in-Aid for Nagoya University Global COE
Program, ``Quest for Fundamental Principles in the Universe: from
Particles to the Solar System and the Cosmos'', and World Premier 
International Research Center Initiative (WPI Initiative), MEXT, Japan.
We acknowledge Kobayashi-Maskawa Institute for the Origin of
Particles and the Universe, Nagoya University for providing computing
resources.
Numerical calculations for the present work have been in part carried out 
under the ``Interdisciplinary Computational Science Program'' in Center for 
Computational Sciences, University of Tsukuba, and also on Cray XT4 at 
Center for Computational Astrophysics, CfCA, of National Astronomical 
Observatory of Japan. 

\bibliographystyle{aa} % style aa.bst
\bibliography{ref3}

\end{document}